\documentclass[12pt]{article}
\usepackage{graphicx}
\usepackage{natbib} 
\usepackage{url} 

\newcommand{\blind}{0}

\usepackage{bm}
\usepackage{mathtools} 
\usepackage{amsmath} 
\usepackage{amssymb}
\usepackage{lineno}

\usepackage{subdepth} 
\usepackage{graphicx}
\usepackage{caption} 
\usepackage{subcaption} 
\usepackage[toc,page]{appendix}
\usepackage{float}
\usepackage{textcomp} 

\usepackage{bookmark}
\bookmarksetup{open,numbered}

\usepackage{tabularx}
\newcolumntype{Y}{>{\centering\arraybackslash}X}
\usepackage{multirow}

\usepackage{booktabs} 
\usepackage[shortlabels]{enumitem} 

\usepackage{algorithm}
\usepackage{algpseudocode}
\usepackage{setspace} 

\begin{document}

\def\spacingset#1{\renewcommand{\baselinestretch}%
{#1}\small\normalsize} \spacingset{1}


\if0\blind
{
  \title{\bf A Framework for Supervised and Unsupervised Segmentation and Classification of Materials Microstructure Images}
  \author{Kungang Zhang\textsuperscript{1} \\
    Wei Chen\textsuperscript{2} \\
    Wing K. Liu\textsuperscript{2} \\
    L. Catherine Brinson\textsuperscript{3} \\
    Daniel W. Apley\textsuperscript{1} \\
    \textsuperscript{1}Department of Industrial Engineering and Management Sciences,\\ Northwestern University \\
    \textsuperscript{2}Department of Mechanical Engineering,\\ Northwestern University \\
    \textsuperscript{3}Department of Mechanical Engineering \& Materials Science,\\ Duke University \\
    }
  \maketitle
} \fi

\if1\blind
{
  \bigskip
  \bigskip
  \bigskip
  \begin{center}
    {\LARGE\bf A Framework for Supervised and Unsupervised Segmentation and Classification of Materials Microstructure Images}
\end{center}
  \medskip
} \fi

\bigskip
\begin{abstract}
       Microstructure of materials is often characterized through image analysis to understand processing-structure-properties linkages. We propose a largely automated framework that integrates unsupervised and supervised learning methods to classify micrographs according to microstructure phase/class and, for multiphase microstructures, segments them into different homogeneous regions. With the advance of manufacturing and imaging techniques, the ultra-high resolution of imaging that reveals the complexity of microstructures and the rapidly increasing quantity of images (i.e., micrographs) enables and necessitates a more powerful and automated framework to extract materials characteristics and knowledge. The framework we propose can be used to gradually build a database of microstructure classes relevant to a particular process or group of materials, which can help in analyzing and discovering/identifying new materials. The framework has three steps: (1) {preliminary} segmentation of multiphase micrographs so that different microstructure homogeneous regions can be identified in an unsupervised manner; (2) {identification and classification of} homogeneous regions of micrographs through an uncertainty-aware supervised classification network trained using the segmented micrographs from Step $1$ with their identified labels verified via the built-in uncertainty quantification and minimal human inspection; (3) {subsequent} supervised segmentation (more powerful than the segmentation in Step $1$) of multiphase microstructures through a segmentation network trained with micrographs and the results from Steps $1$-$2$ using a form of data augmentation. This framework can iteratively characterize/segment new homogeneous or multiphase materials while expanding the database to enhance performance. The framework is demonstrated on various sets of materials and texture images.
\end{abstract}

\noindent%
{\it Keywords:}  Microstructure, Fisher Score Vector, Evidential Deep Learning, Segmentation Network, Data Augmentation
\vfill

\newpage
\spacingset{2} 
\section{Motivation and Introduction}
\label{s:motiv}
\begin{figure}[!htbp]
\centering
       \begin{subfigure}[t]{0.42\linewidth}
              \centering
              \includegraphics[width=\textwidth, trim=.0in .0in .0in .0in, clip]{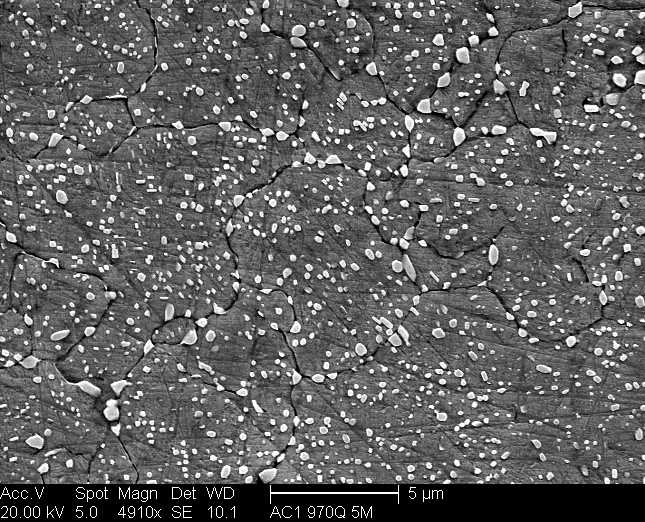}
              \captionsetup{width=.95\linewidth}
              \caption{An SEM image of UHCS with single spheroidite phase.}
              \label{fig_DL:uhcs_singlephase}
       \end{subfigure}
       \begin{subfigure}[t]{0.42\linewidth}
              \centering
              \includegraphics[width=\textwidth, trim=.0in .0in .0in .0in, clip]{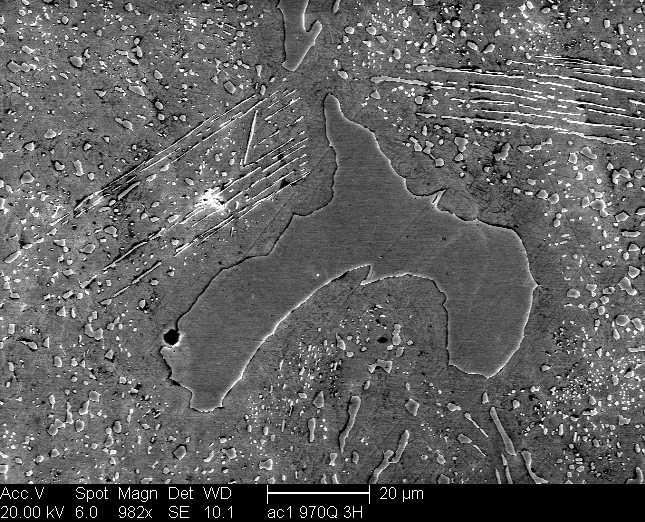}
              \captionsetup{width=.95\linewidth}
              \caption{An SEM micrograph of UHCS with multiphase: ferritic, proeutectoid, spheroidite, and Widmanst\"atten.}
              \label{fig_DL:uhcs_multiphase}
       \end{subfigure}
       \begin{subfigure}[t]{0.182\linewidth}
              \centering
              \includegraphics[width=\textwidth, trim=.0in .0in .0in .0in, clip]{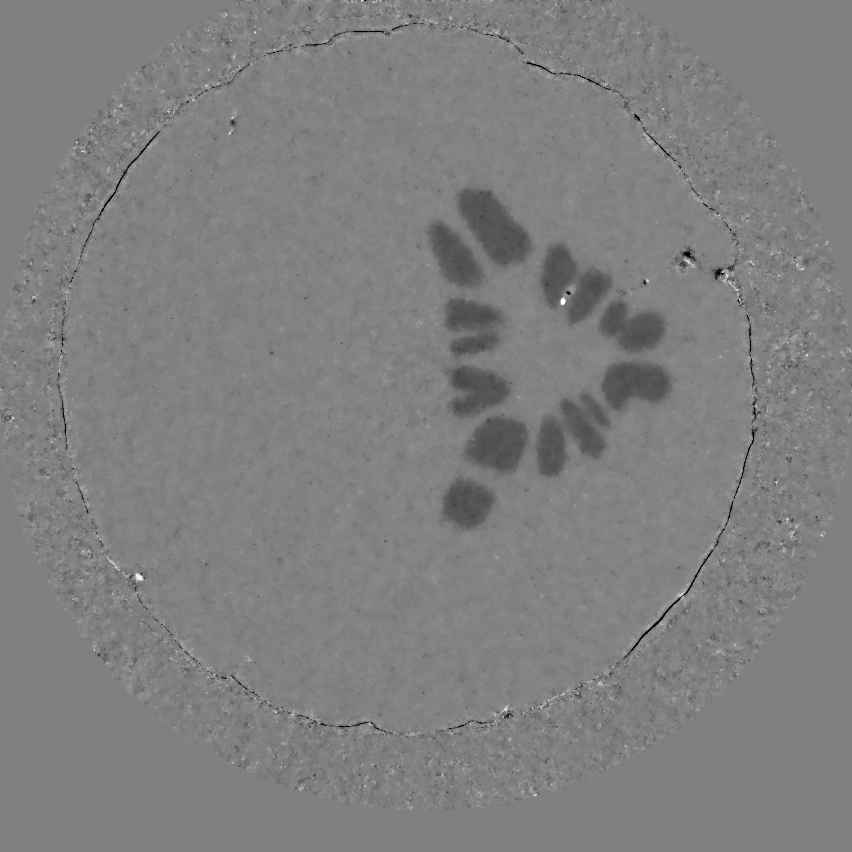}
              \captionsetup{width=.95\linewidth}
              \caption{Al-Zn alloy.}
              \label{fig_DL:xct}
       \end{subfigure}
       \begin{subfigure}[t]{0.3118\linewidth}
              \centering
              \includegraphics[width=\textwidth, trim=.0in .0in .0in .0in, clip]{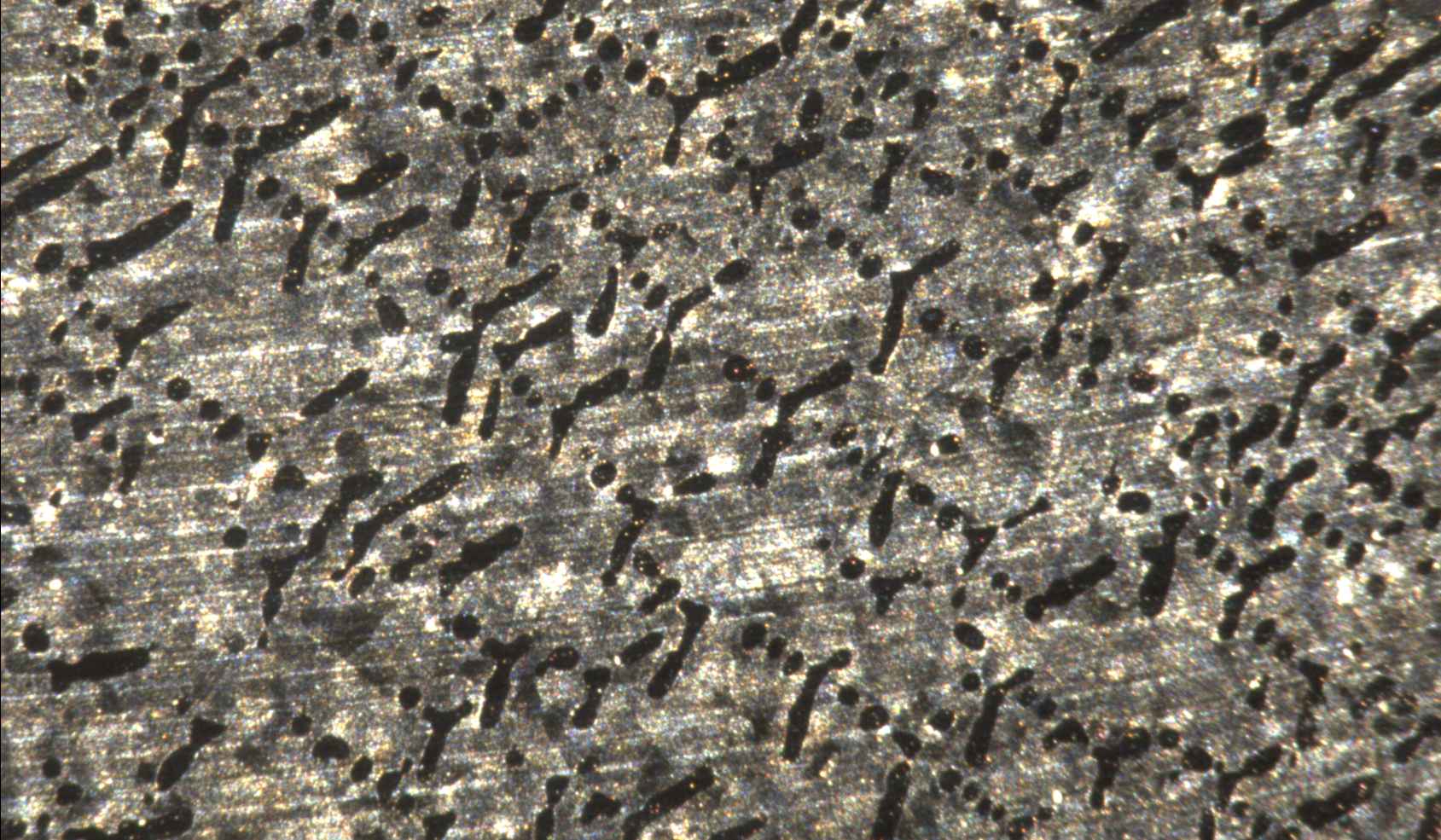}
              \captionsetup{width=.95\linewidth}
              \caption{Pb-Sn alloy.}
              \label{fig_DL:ss}
       \end{subfigure}
       \begin{subfigure}[t]{0.2235\linewidth}
              \centering
              \includegraphics[width=\textwidth, trim=2.1in .6in 7.in 5.in, clip]{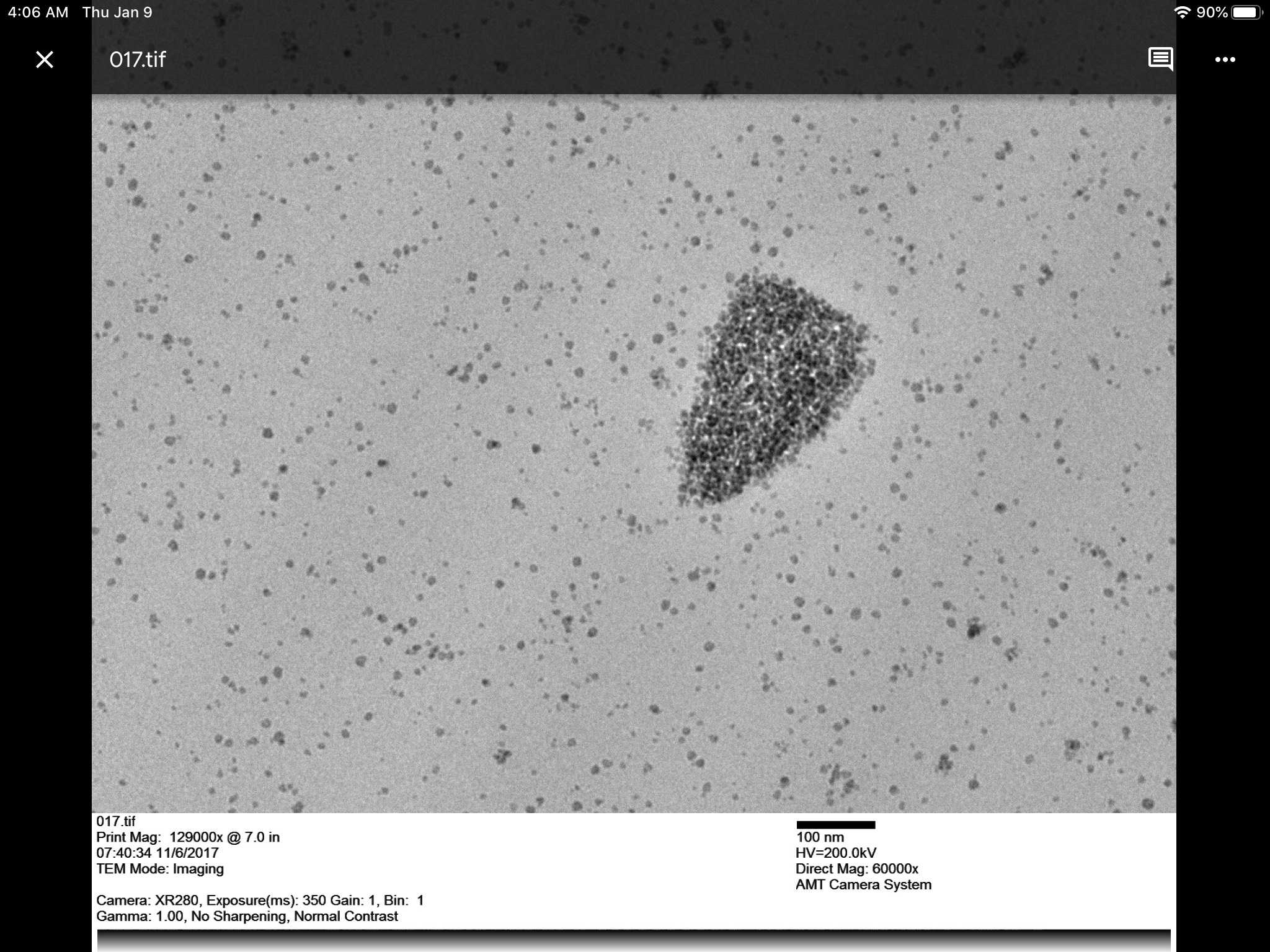}
              \captionsetup{width=.95\linewidth}
              \caption{Silica in PMMA.}
              \label{fig_DL:silica_PMMA}
       \end{subfigure}
       \begin{subfigure}[t]{0.242\linewidth}
              \centering
              \includegraphics[width=\textwidth, trim=.0in .0in .0in .0in, clip]{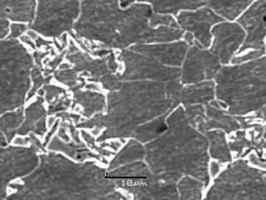}
              \captionsetup{width=.95\linewidth}
              \caption{Dual-phase steel.}
              \label{fig_DL:dual_phase}
       \end{subfigure}
       \addtocounter{footnote}{-1}
       \caption[Caption for LOF]{Examples of single phase (a) and multiphase materials (b)-(f){\protect\footnotemark{}}. (a)-(b) Two SEM images of the UHCS data set~\cite{decost2017uhcsdb,hecht2016digital}. Notice that the two images are not in the same magnification, a commonly {encountered} issue of data quality in real data sets. (c) A X-ray CT~(XCT) image of dendrite growth from a solidification experiment on {a }Al-Zn alloy~\cite{stan2020optimizing}. (d) A serial sectioning~(SS) image of dendrite growth from a coarsening experiment on {a }Pb-Sn alloy~\cite{stan2020optimizing}. (e) A TEM image of silica particles in PMMA. (f) An SEM image of dual-phase steel~\cite{banerjee2013segmentation}.}
       \label{fig_DL:mat_example}
\end{figure}

Microstructure plays a vital role in determining physical properties of materials and understanding the relations among the processes, structures, and properties~\cite{decost2015computer}. Recently, analyzing images of multiphase materials, often called micrographs~\cite{o2016materials,puchala2016materials,michel2016beyond}, has attracted much attention across different materials fields~(e.g., metals~\cite{lewandowski2016metal}, polymer composites~\cite{ligon2017polymers}, and ceramics~\cite{chen20193d,moritz2018additive}, etc.) due to burgeoning manufacturing and imaging techniques~\cite{balla2012laser, bandyopadhyay2018additive}. For example, the ultra-high carbon steel (UHCS) data set~\cite{decost2017uhcsdb} includes $961$ SEM micrographs, of which two multiphase microstructures are shown in Figs.~\ref{fig_DL:uhcs_singlephase} and~\ref{fig_DL:uhcs_multiphase}. Fig.~\ref{fig_DL:uhcs_singlephase} is a micrograph containing a single spheroidite phase, while Fig.~\ref{fig_DL:uhcs_multiphase} contains ferritic, proeutectoid, spheroidite, and Widmanst\"atten phases. Because the phase characteristics impact material properties (e.g., toughness {and crack initiation etc.})~\cite{hecht2016digital,carter2014influence}, identifying and classifying the individual phases present in a micrograph is an important step towards understanding the physical properties of the material. Throughout this paper, by ``multiphase'' we mean a microstructure whose stochastic nature (different phases have different stochastic nature) varies spatially, sometimes referred to as microstructure nonstationarity. This includes certain steels that have different phases present simultaneously by design, as well as certain composites whose stochastic microstructure composition varies spatially due to variation in the processing conditions. By ``single-phase'', we mean a homogeneous microstructure whose stochastic nature is spatially stationary. Throughout the paper, the term ``microstructure class'' (MC) refers to the label assigned to each homogeneous region (HR) of connected pixels within a micrograph with each HR comprised of a single phase.

\footnotetext{Original micrographs with ultra-high resolution can be found through links in~\cite{decost2017uhcsdb}.}
The primary purpose of this paper is to introduce a framework for segmenting each micrograph into HRs/phases and classifying each HR as either an existing MC from a database of previously-identified MCs or a new MC. As new MCs are identified in our framework, they are added to the database. This framework is versatile in a sense that it can be applied to micrographs collected by different instruments~(e.g., SEM, TEM, X-ray CT, Optical Microscopy, etc.~\cite{puchala2016materials,gagliardi2015material,takahashi2016materials,blaiszik2016materials,jain2016new,jain2016research,kim2016organized,rose2017aflux,AFRL2018}) for various materials~(e.g., metals~\cite{lewandowski2016metal}, polymer composites~\cite{ligon2017polymers}, and ceramics~\cite{chen20193d,moritz2018additive}, etc.) across many industrial applications~(e.g., additive manufacturing~\cite{balla2012laser, bandyopadhyay2018additive}, quality control~\cite{bui2018monitoring}).

Characterizing material microstructures has long been a major research focus. Recently, machine learning (ML), deep learning (DL), and AI have received growing interest in this area~\cite{stan2020optimizing,maik2020big,ajioka2020development,agrawal2019deep}. These approaches transform traditional methods of characterization, enabling data-driven Processing-Structure-Properties-Performance (PSPP) linkages~\cite{agrawal2016perspective}. Traditionally, methods for pre-processing~(for many downstream tasks) and analyzing multiphase micrographs were largely manual or semi-automated {and time consuming. For example, one week of synchrotron beamtime can produce $\sim 2.3$ millions micrographs, one of which is shown in Fig.~\ref{fig_DL:xct}, and manual segmentation of different metal phases of one micrograph takes a researcher $\sim 20$ minutes, which translates to $\sim 88$ years of human involvement to manually process one-week worth of data~\cite{stan2020optimizing}.} Another drawback of traditional methods is that they are not generic to materials and fail to fully represent and distinguish complex morphologies~\cite{kondo2017microstructure}. For example, some methods rely on handcrafted descriptors (e.g., volume fraction, dispersion)~\cite{ashby1993materials,xu2014descriptor}. Statistical encoders (e.g., N-point correlation functions {(NPCF)}) can better represent microstructures but are often computationally expensive, depend on prior knowledge of patch size, and are limited in characterizing complex { and stochastic} patterns in generic materials microstructures~\cite{decost2015computer,yabansu2017extraction,xu2015machine,yu2017characterization, zhang2021nonstationarity}. The rich micrograph data sources that are now available warrant more automated methods that take advantage of modern machine learning algorithms. ML/DL can learn features directly from raw images. For instance, convolutional neural networks (CNNs) eliminate manual feature extraction and have achieved high accuracy in segmentation, phase identification, and particle size estimation~\cite{decost2019high,jang2019residual,decost2017exploring}. Even though many machine learning methods have been effectively applied to materials science problems, including micrograph segmentation, there remain challenges in expanding this interdisciplinary research: Much of the existing work centers on supervised learning that requires a large amount of labeled data (e.g., micrographs labeled pixel-by-pixel according to their microstructure class)~\cite{kondo2017microstructure,lubbers2017inferring,azimi2018advanced,strohmann2019semantic,chen2019aluminum} to train a segmentation model to reach a high accuracy. Moreover, works utilizing unsupervised or semi-supervised learning methods usually do not require labels during training~\cite{liu2024review, zhang2023unleashing, huang2023self} but still need manual identification and/or classification and are more expensive and less powerful than their supervised counterparts in production~\cite{zhang2021nonstationarity, decost2017exploring,impoco2015incremental,kitahara2018microstructure}. This study proposes a hybrid framework combining aspects of both approaches. By leveraging minimal labels and iterative refinement, we aim to produce accurate, scalable, and efficient microstructure characterization. 

{A main contribution of this paper is the overall framework that integrates various unsupervised learning and supervised learning concepts to initially segment a micrograph into HRs (using unsupervised learning), classify the identified HRs, and then refine the segmentation using a supervised segmentation approach. Applied iteratively to a series of micrographs, the framework can be used to build a database of stochastic microstructures that is incorporated into the supervised segmentation and MC classification steps. The unsupervised learning Step $1$ is based on the approach of~\cite{zhang2021nonstationarity} but uses Bayesian Gaussian Mixture clustering to more systematically determine the number of HRs. The classification Step $2$ is novel in the sense that it is first informed by the unsupervised segmentation from Step $1$ so that each identified HR from Step $1$ is classified as a single entity (as opposed to classifying each pixel or small region of pixels in the original micrograph), which is more efficient since it allows information over the entire HR to be used for classification. We also incorporate evidential deep learning into Step $2$, which allows the framework to assess the likelihood that a newly collected HR belongs to a new MC that is not currently in the database. The supervised segmentation in Step $3$ incorporates a novel form of transfer learning using a database of stochastic texture images (of various fabrics, grass, seeds, ceiling textures, etc) to pretrain the model. Although these pretraining images are not of material microstructures, we demonstrate that it substantially improves the material microstructure segmentation and classification and mitigates the problem of collecting labelled segmentation data of materials microstructures.} 

{The format of the remainder of the paper is as follows. }In Section~\ref{s_DL:model_ms}, we discuss the theoretical foundations and methods for each step of our framework. In Section~\ref{s_DL:data_set_exp}, data sets, demonstrations, and experiment results for each step are presented, demonstrating that our approach can enhance characterization tasks with minimal human intervention.

\section{Theoretical Foundation and Methodology}
\label{s_DL:model_ms}
To address the motivations and challenges discussed in the introduction, we propose a framework for characterizing single-phase or multiphase microstructures with the following major steps and advantages~(also see the flow chart in Fig.~\ref{fig_DL:flow_chart_img_analy}):

\begin{figure}[!htbp]
\centering
\includegraphics[width = 1\linewidth, trim=1.6in .2in 1.6in .25in, clip]{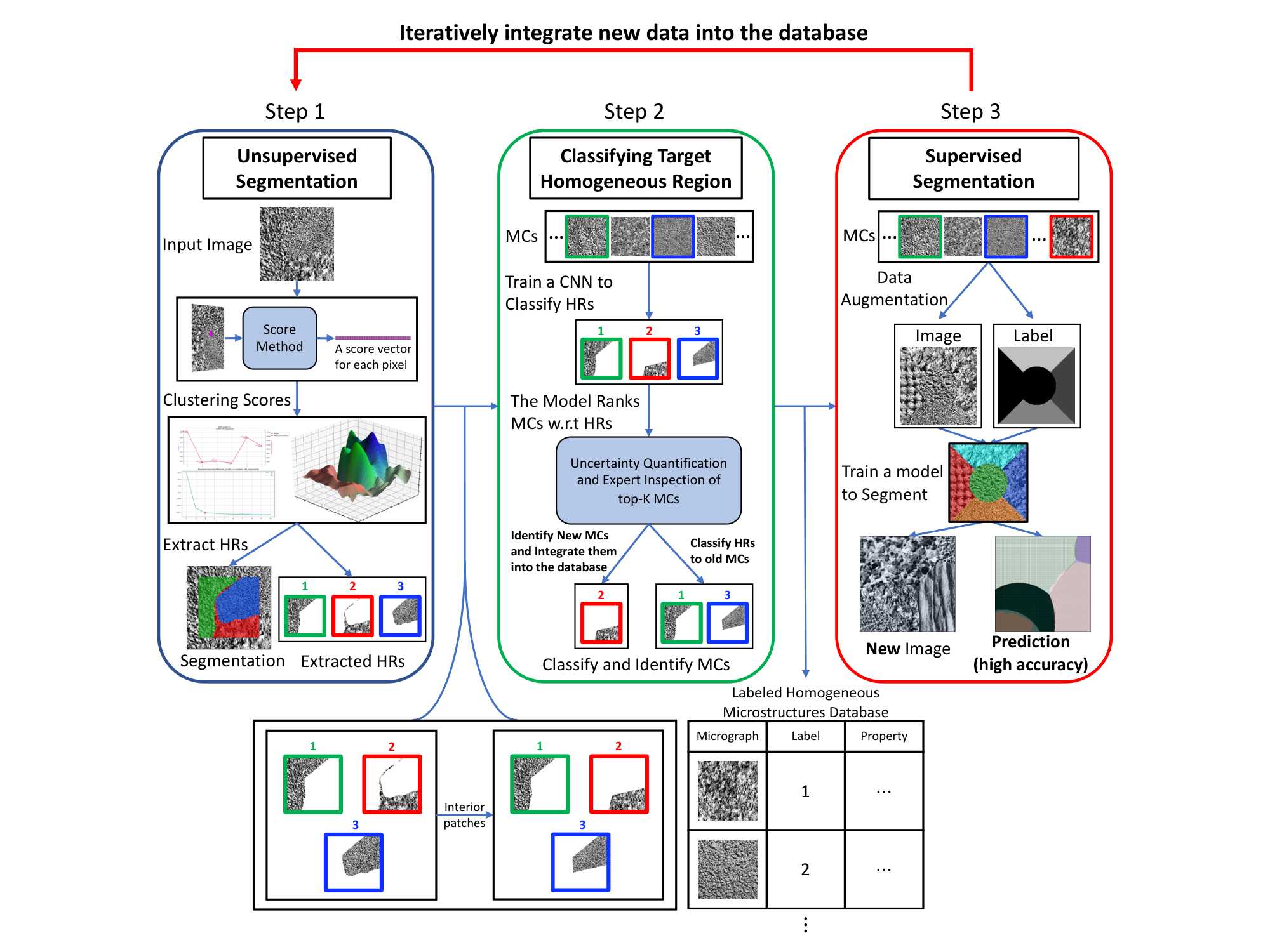}
\caption{Flow chart of the iterative framework. Between steps, the knowledge passed from the previous to the next are shown: from Step $1$ to $2$, the HRs are extracted and entered into the database of existing MCs; from Step $2$ to $3$, the database of classified and labeled homogeneous microstructures is passed for data augmentation and training segmentation networks.}
\label{fig_DL:flow_chart_img_analy}
\end{figure}

\textbf{Step $1$}: \textit{In an unsupervised manner, characterize multiphase nonstationary behavior in micrographs using a recently developed score-based nonstationarity analysis method~\cite{zhang2021nonstationarity}~(see Sec.~\ref{ss_DL:nd_score}). Specifically, we fit a single supervised learning model to a set of training micrograph(s) to predict the grayscale value for each pixel, as a function of some window of neighboring pixels, which provides a fingerprint to the stochastic nature of the microstructure. We then apply the model to predict each pixel of the training micrograph(s) to obtain score vectors (see Eq.~(\ref{eqn_DL:score_func}) later) pixel-by-pixel, and then cluster score vectors to segment HRs corresponding to distinct MCs. Note that our use of supervised learning in this step differs from existing supervised learning approaches in which the supervised learning model is a classification model to classify MC at each pixel, which requires MC labels to train.}

\textbf{Step $2$}: \textit{The extracted HRs from Step $1$ are either classified as one of the previously identified and analyzed MCs in the database or{, based on our classification uncertainty quantification,} determined to be a new MC not previously cataloged. To do this, an uncertainty-aware model predicts which existing MC a target HR belongs to and ranks the different MCs based on similarity to the target HR along with an uncertainty measure of the confidence of the prediction. If the target HR has high uncertainty it is concluded to be different from any existing MC (which can be verified with minimal human inspection), and the new MC is integrated into the database and the classification model. This largely shifts the burden of screening/memorizing morphologies and patterns of an entire database of MCs from humans to algorithms and thus accelerates discovery/identification of new materials.}

\textbf{Step $3$}: \textit{The database of MCs and their corresponding labels are then used to train a supervised segmentation model through data augmentation to improve segmentation quality relative to {the segmentation in }Step $1$. This provides more automated, more accurate, and faster identification, classification, and segmentation of MCs for potential use for in-situ monitoring and real-time control of production{/processing}.}

\textbf{Iteration over Steps $1$-$3$}: \textit{As new micrographs are collected, they pass through the above three steps to classify/segment their phases and label any new, previously unseen, phases as new MCs (as in Step $2$) and to provide additional samples of existing MCs. The iterations are intended to improve {the }data-driven models~(e.g., classification or segmentation neural networks) by either increasing the number of MCs they can classify (if one of the phases is labeled as a new MC) or by further training the models on the newly acquired additional samples of existing MCs.}

As described in the Introduction, our iterative framework integrates supervised and unsupervised learning in a way that combines the strengths of these two types of existing approaches for micrograph segmentation and classification. In the remainder of this section, we describe the details of each step of the framework.

\subsection{Step 1: Micrograph nonstationarity analysis and unsupervised segmentation}
\label{ss_DL:nd_score}
\begin{figure}[H]
       \centering
          \begin{subfigure}[t]{.3\linewidth} 
                \centering
                \includegraphics[width = \linewidth, trim=0in 0.05in 0in 0in, clip]{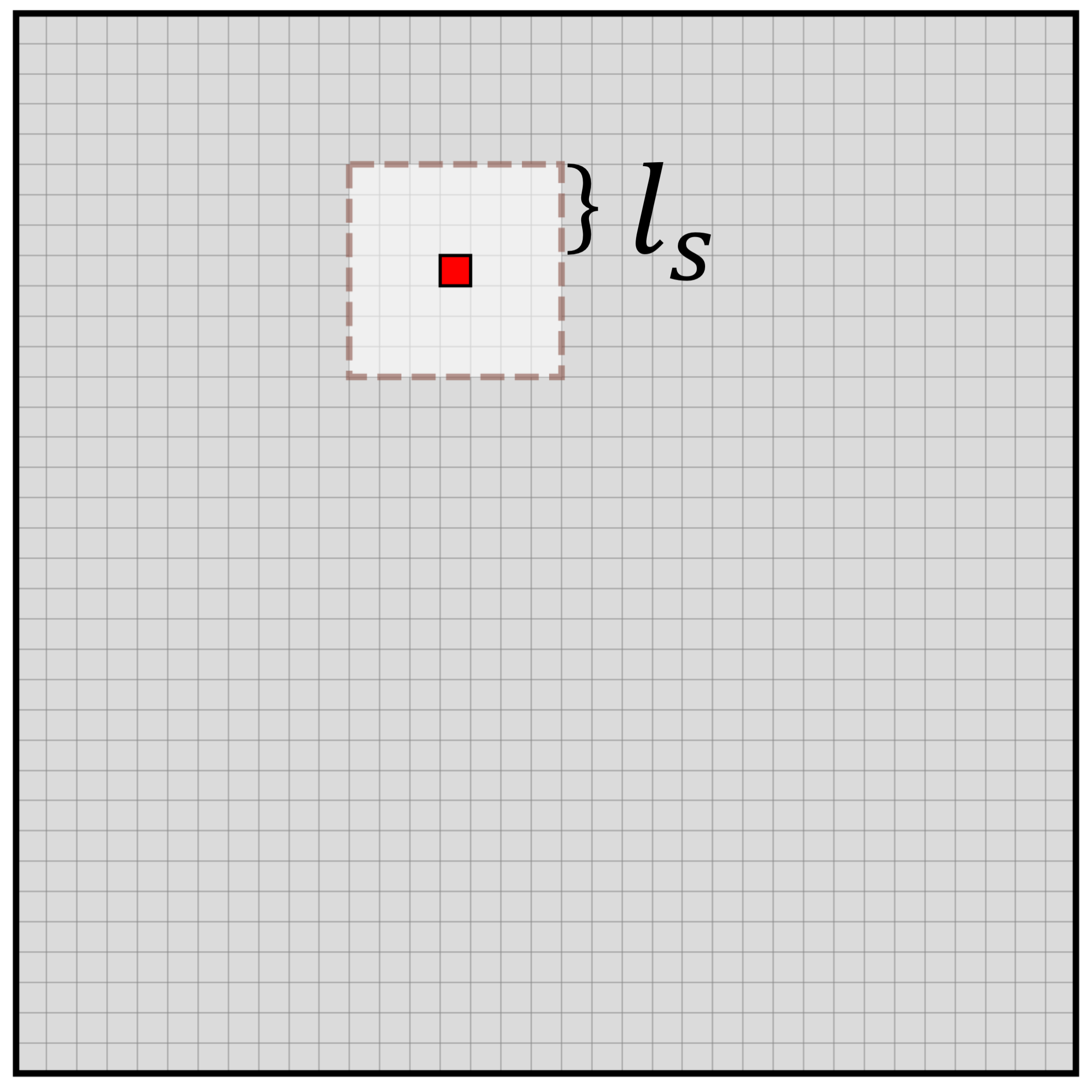}
                \captionsetup{width=.95\linewidth}
              \caption{Non-causal neighborhood window.}
                \label{fig_DL:non_causal_wind}
         \end{subfigure}
         \begin{subfigure}[t]{.3\linewidth} 
                \centering
                \includegraphics[width = \linewidth, trim=0in 0in 0in 0in, clip]{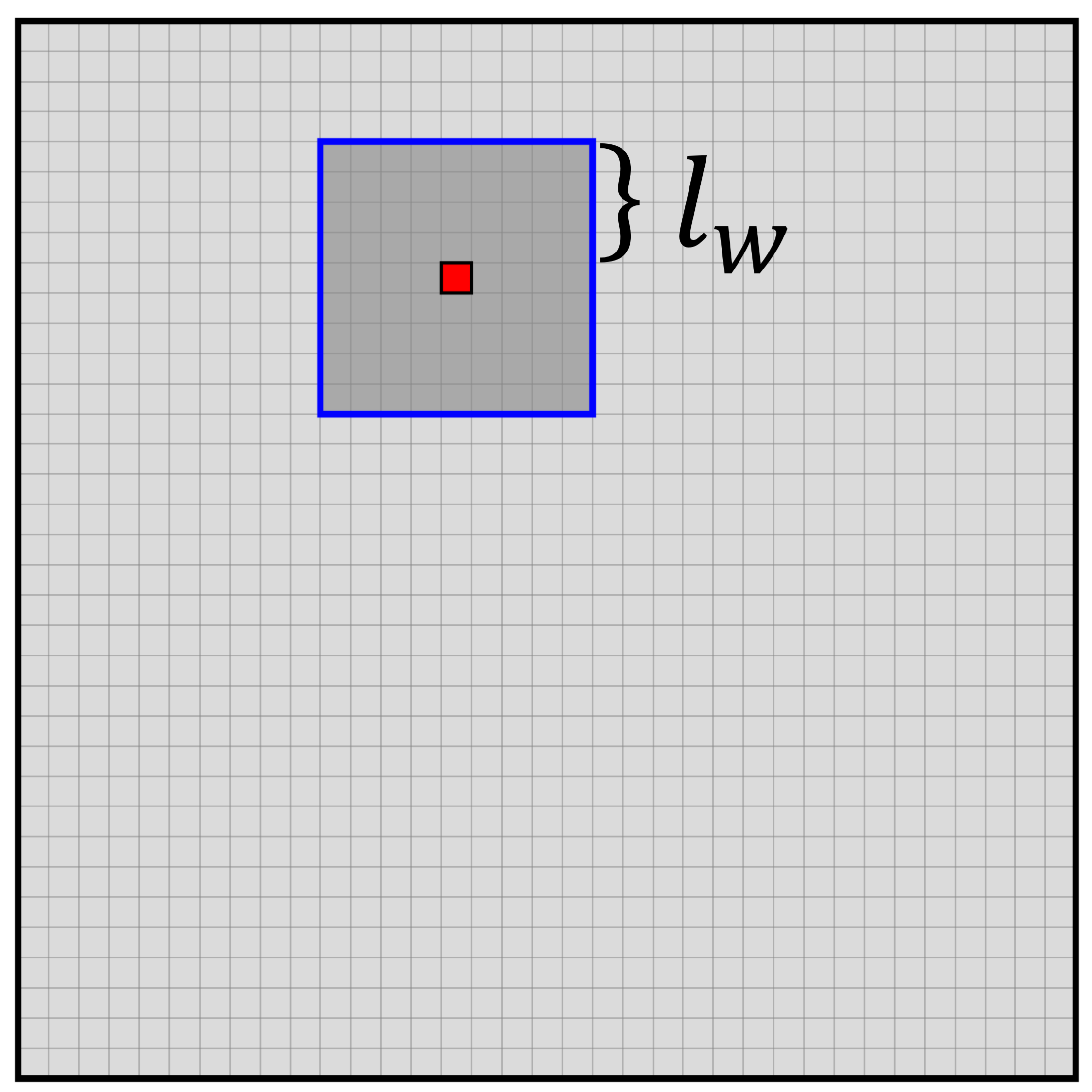}
                \captionsetup{width=.95\linewidth}
              \caption{Spatial WMA window.}
                \label{fig_DL:sewma_wind}
         \end{subfigure}
       \begin{subfigure}[t]{.36\linewidth} 
                \centering
                \includegraphics[width = \linewidth, trim=0.in -1.0in .0in .0in, clip]{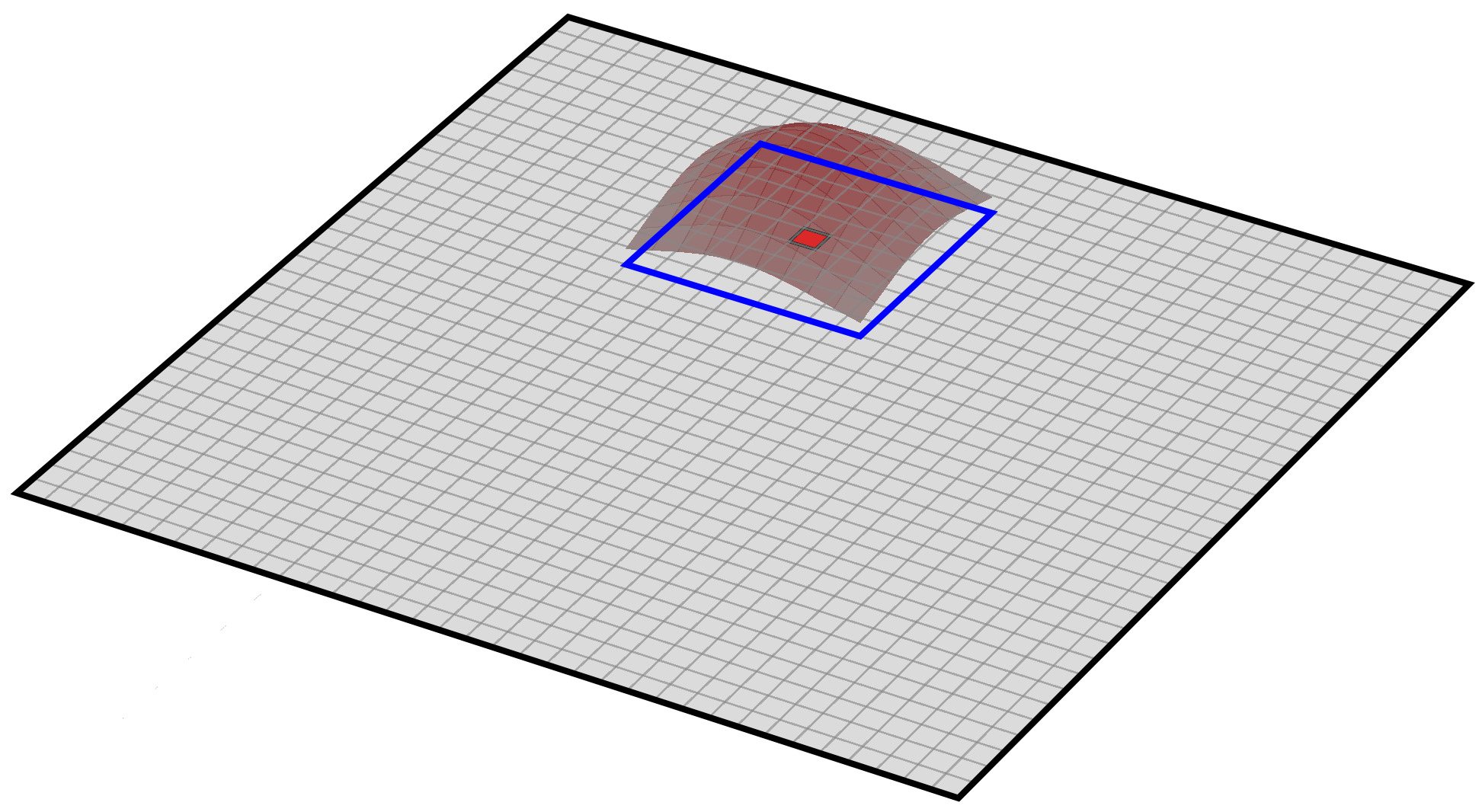}
                \captionsetup{width=.95\linewidth}
              \caption{Gaussian distribution over the spatial WMA window.}
                \label{fig_DL:gau_dist_sewma}
         \end{subfigure}
       \caption{Illustration of processing score vectors. (a) Neighborhood window (brown dashed square excluding the target pixel colored as red) for modeling the conditional distribution of $X_i$ (the red pixel) given the neighboring pixels in the window. (b) Spatial WMA window (blue square including the target pixel colored as red) for spatially smoothing score vectors. (c) A $2$D Gaussian weighting function is centered (at the red pixel) and truncated over the WMA window. The height of profile and grayscale value is proportional to the density value of the weighting function.}
       \label{fig_DL:causal_non_causal_wind}
\end{figure}

In general, the microstructure of materials can be modeled as realizations of a spatial random process in the following manner. Denote the concatenated $m$ pixel values in a micrograph as a vector $\bm{X} = [X_1, X_2, \cdots, X_m]$. The value at the $i^{\mathrm{th}}$ pixel, conditioned on all other pixels in the micrograph, can be seen as a random variable $X_i$ drawn from a conditional distribution that is approximated as $P({X}_i|{\bm{\mathbb{N}}}(X_i))$, where ${\bm{\mathbb{N}}}(X_i)$ denotes some local neighborhood of pixel values surrounding the $i ^{\mathrm{th}}$ pixel as shown in Fig.~\ref{fig_DL:non_causal_wind}~\cite{bostanabad2016stochastic}. Via Gibbs sampling arguments, this conditional distribution $P({X}_i|{\bm{\mathbb{N}}}(X_i))$ can be viewed as characterizing the joint distribution of all pixels in the micrograph and, thus, the stochastic nature of the microstructure. 

Any appropriate supervised learning model (e.g., a neural network) can be fit to all the observations $\{(X_i, \bm{\mathbb{N}}({X}_{i}))\}_{i=1}^{m}$ (with $X_i$ and $\bm{\mathbb{N}}({X}_{i})$ being the response and the predictor 
variables, respectively) from a training micrograph(s), as a means of modeling the distribution $P({X}|\bm{\mathbb{N}}(X); \bm { \theta})$ with {model }parameters denoted by $\bm{\theta}$ (e.g., the weights for all neurons in a neural network). $\bm{\theta}$ implicitly represents the distribution of the pixels in the micrograph and therefore provides a relatively compact ``fingerprint'' that represents the stochastic nature of the microstructure. We adopt the approach developed in \cite{zhang2021nonstationarity} for monitoring for nonstationarity in the nature of the microstructure using Fisher score vectors {(which was based on the more general concept drift detection approach of \cite{zhang2023concept})}, which, for the $i ^{\mathrm{th}}$ pixel{, is defined} as
\begin{align}
\bm {s}(\bm {\theta}; y_i, \bm{x}_i) = \nabla_{\bm {\theta}} \log P(X = y_i | \bm{\mathbb{N}}(X) = \bm{x}_i; \bm {\theta})
\label{eqn_DL:score_func}
\end{align}
where $\nabla_{\bm {\theta}}$ is the gradient operator with respect to $\bm {\theta}${; $y_i$ and $\bm{x}_i$ are the observations of the $i^{\mathrm{th}}$ pixel and the vector of pixels over its chosen local neighborhood}. For grayscale images, the conditional distribution of $X$ given $\bm{\mathbb{N}}(X)=\bm {x}$ is approximated as normal with mean $g(\bm{x};\bm{\theta})$ and variance $\sigma^2$, where $g(\bm{x};\bm{\theta})$ is the prediction of $X$ from the supervised learning model. Under mild conditions, if the microstructure is stationary over the region associated with the training data and the supervised learning model is ``correct'', the score vector is zero-mean in the sense that
\begin{align}
E_{\bm {\theta}} [\bm {s}(\bm {\theta}; X, \bm{\mathbb{N}}(X))] = \bm {0}
\label{eqn_DL:score_mean_zero_theory}
\end{align}
where $E_{\bm{\theta}}[\cdot]$ denotes the expectation operator (with respect to the random variables $X$ and $\bm{\mathbb{N}}(X)$), and the subscript $\bm { \theta}$ indicates that the distribution of pixel value $X$ and its neighborhood values $\bm{\mathbb{N}}(X)$ are based on the supervised learning model with true parameters $\bm { \theta}$. Conversely, if the local stochastic nature in the vicinity of pixel $X$ changes from what it was over the training data~(i.e., the true parameters change from $\bm{\theta}$ to $\bm { \theta}' \neq \bm { \theta}$ at $X$), then the mean of the score vector at $X$ for a model with parameters $\bm{\theta}$ generally differs from zero~(i.e., $E_{\bm {\theta}'} [\bm {s}(\bm {\theta}; X, \bm{\mathbb{N}}(X))] \neq \bm {0}$). In practice, $\bm{\theta}$ for the training data is replaced by its maximum likelihood estimate $\hat{\bm{\theta}}$, and the expectation in Eq.~(\ref{eqn_DL:score_mean_zero_theory}) is replaced by the empirical mean $\hat{E}_{\bm { \theta}}[\bm{s}(\hat{\bm{\theta}};X, \bm{\mathbb{N}}(X))]${ over all pixels in the micrograph training data}~(for details refer to~\cite{zhang2021nonstationarity}). Here, the subscript $\bm { \theta}$ indicates whatever parameters are valid in the vicinity of pixel $X$, and $\hat{\bm{\theta}}$ indicates the {globally }estimated parameters over the training data. {The estimate $\hat{\bm{\theta}}$ fit to the entire micrograph can be thought of as an aggregation of the $\bm{\theta}$ for each MC in the micrograph, and the average of the score vectors $\bm{s} (\hat{\bm{\theta}}; y_i, \bm {x}_i)$ over the entire micrograph is the zero vector.}

From the preceding {and the following arguments, which summarize the theory in~\cite{zhang2021nonstationarity}}, the problem of analyzing nonstationarity in the microstructure over a micrograph(s) is converted to fitting a single supervised learning model to the micrograph(s), applying the model to predict each pixel, and then analyzing whether the mean score vector locally differs substantially from the zero vector anywhere in the micrograph(s). If the stochastic nature of different MCs, as captured by $\bm{\theta}$, is distinct enough, the score vectors $\bm{s} (\hat{\bm{\theta}}; y_i, \bm {x}_i)$ belonging to different HRs should center around different mean vectors and be separable in the score-vector space. Thus, a clustering algorithm can be applied to the (smoothed version of, via some spatial moving window) score vectors to diagnose the spatial nonstationarity of micrographs or, more specifically, segment different HRs based on their stochastic natures in an unsupervised manner \cite{zhang2021nonstationarity}. To mitigate the effects of noise in the score vectors on the clustering performance, before clustering, {we calculate }a spatial Gaussian weighted moving average (WMA) of score vectors, denoted as $\bm {z}_i$ for the $i ^{\mathrm{th}}$ pixel, where the WMA window and the $2$D weight density functions are depicted in Figs.~\ref{fig_DL:sewma_wind} and~\ref{fig_DL:gau_dist_sewma}. 

To cluster score vectors and segment the micrograph into HRs, we use the Bayesian Gaussian Mixture (BGM) clustering method~\cite{bishop2006pattern}. Within the clustering algorithm, the number of clusters (corresponding to the number of HRs) must be estimated. For this we use AIC/BIC curves and posterior weights plot produced within the BGM clustering algorithm. We focus on the BGM approach, which is a more computationally expensive clustering method than the basic \textit{k-means} clustering method, because it provides more information on the number of clusters. More specifically, we use the following Bayesian Gaussian mixture model for the smoothed version of the score vectors $\bm{z}_i$ with their latent class one-hot encoding labels $\bm{c}_i$:

\begin{align}
&p(\{\bm{z}_i\}_{i=1}^{m}, \{\bm{c}_i\}_{i=1}^{m}, \bm{\pi}, \bm{\mu}, \bm{\Lambda}) \nonumber \\
=& \prod_{i=1}^{m} p(\bm{z}_i | \bm{c}_i, \bm{\mu}, \bm{\Lambda}) p(\bm{c}_i | \bm{\pi}) p(\bm{\pi}) p(\bm{\mu}, \bm{\Lambda}) \nonumber \\
=& \prod_{i=1}^{m} \prod_{k=1}^{K} \mathcal{N}(\bm{z}_i| \bm{\mu}_k, \bm{\Lambda}_k^{-1})^{c_{i,k}} \pi_k^{c_{i,k}} p(\bm{\pi}) p(\bm{\mu}, \bm{\Lambda})
\label{eqn_DL:bgm_posterior_form}
\end{align}
where $\bm{\pi}=[\pi_1, \pi_2,\cdots,\pi_K]$ is the mixing proportion of each cluster; $\bm{\mu}=\{\bm{\mu}_k\}_{k=1}^K$ and $\bm{\Lambda}=\{\bm{\Lambda}_k\}_{k=1}^K$ are the mean vectors and precision matrices of Gaussian distribution associated with each cluster, respectively; and $\bm{c}_i=[c_{i,1},c_{i,2},\cdots\\,c_{i,K}]$ is the standard one-hot encoding vector of zeros with a one in the $k^{\mathrm{th}}$ position if $\bm{z}_i$ is in cluster $k$. The last equation above utilizes the relations $p(\bm{z}_i | \bm{c}_i, \bm{\mu}, \bm{\Lambda}) = \prod_{k=1}^{K} \mathcal{N}(\bm{z}_i| \bm{\mu}_k, \bm{\Lambda}_k^{-1})^{c_{i,k}}$ and $p(\bm{c}_i | \bm{\pi})=\prod_{k=1}^{K}\pi_k^{c_{i,k}}$. The prior distributions of parameters $\bm{\pi}$, $\bm{\mu}$, and $\bm{\Lambda}$ are:
\begin{align}
p(\bm{\pi}) &= \text{Dir}(\bm{\pi}|\bm{\alpha}_0) \\
p(\bm{\mu}, \bm{\Lambda}) &= \prod_{k=1}^{K} \mathcal{N}(\bm{\mu}_k | \bm{\mu}_0, (\beta_0 \bm{\Lambda}_k)^{-1}) \text{Wishart}(\bm{\Lambda}_k | \bm{W}_0, \nu_0)
\label{eqn_DL:bgm_prior}
\end{align}
where $\text{Dir}(\bm{\pi}|\bm{\alpha}_0)$ is the Dirichlet distribution with parameter vector $\bm{\alpha}_0$; \\$\mathcal{N}(\bm{\mu}_k | \bm{\mu}_0, (\beta_0 \bm{\Lambda}_k)^{-1})$ is the multivariate Gaussian distribution with mean vector $\bm{\mu}_0$ and precision matrix $\beta_0 \bm{\Lambda}_k$; and $\text{Wishart}(\bm{\Lambda}_k | \bm{W}_0, \nu_0)$ is the Wishart distribution with scale matrix $\bm{W}_0$ and degrees of freedom $\nu_0$. We follow standard rules~\cite{bishop2006pattern,blei2003latent,teh2004sharing} for choice of all hyperparameters $\bm{\alpha}_0$, $\bm{\mu}_0$, $\beta_0$, $\bm{W}_0$, and $\nu_0$. In particular, we use $\alpha_{0,k}=10^{-9} (k=1,2,\cdots,K)$, $\bm{\mu}_0=\frac{1}{m}\sum_{i=1}^{m}\bm{z}_i$, $\beta_0=1$, $\bm{W}_0^{-1}=\frac{1}{m-1}\sum_{i=1}^{m}(\bm{z}_i-\bar{\bm{z}})(\bm{z}_i-\bar{\bm{z}})^T$, and $\nu_0=dim(\bm{z}_i)$. We then use the standard variational inference algorithm to compute posterior distributions for parameters $\bm{\pi}$, $\{\bm{\mu}_k\}_{k=1}^K$, and $\{\bm{\Lambda}_k\}_{k=1}^K$ and latent variables $\{\bm{c}_i\}_{i=1}^{m}$ and use their maximum a posteriori estimators as point estimates. The posterior estimate of the mixing proportion $\bm{\pi}$ can be used to determine the number of clusters in the micrograph, because the $k^{\mathrm{th}}$ element $\pi_k$ can be thought of as the portion of the effective number of $\bm{z}_i$'s assigned to the $k^{\mathrm{th}}$ cluster. This is a key feature of BGM that can help determine the number of clusters in a data-driven manner. See the later examples (e.g., Figs.~\ref{fig_DL:score_unsupervised_bgm_posterior} and~\ref{fig_DL:score_unsupervised_bgm_aic_bic}) for use of BGM posterior weights and AIC/BIC to determine the number of clusters. After estimating the number of clusters and clustering the score vectors based on the posterior distribution of $\{\bm{c}_i\}_{i=1}^{m}$, (each cluster represents a different MC) HRs can be obtained, possibly with minimal guidance from human experts to verify the unsupervised segmentation of HRs.

{Note that over a region near the micrograph borders with width $l_s+l_w$ ($l_s$ and $l_w$ are the two window half-lengths in Figs.~\ref{fig_DL:non_causal_wind} and~\ref{fig_DL:sewma_wind}), the pixels within this border region have no smoothed score vectors due to the lack of neighboring pixels and therefore cannot be segmented{ in Step $1$}~(Fig.~\ref{fig_DL:score_unsupervised}). Step $3$~(Sec.~\ref{ss_DL:sup_seg}) avoids this limitation and can classify the MC in the region near borders by training a supervised segmentation model incorporating the MCs and verified labels from Steps $1$-$2$.}

\subsection{Step $2$: Classification and ranking of MCs in the database using a CNN}
\label{ss_DL:cla_retri}
After {the }unsupervised segmentation of Step 1, square patches of certain dimension cut from each of the resulting HRs (one corresponding to each identified cluster/segment) is input to a classification model to either determine which MC from the database it corresponds to or to classify it as a new MC to be added to the database. For this we train a form of CNN to automatically classify HRs according to their MCs and identify novel MCs that have not been characterized before.

\begin{figure}[!htbp]
       \centering
        \begin{subfigure}[t]{0.35\linewidth}
              \centering
              \includegraphics[width = \linewidth, trim=0in 0in 0in 0in, clip]{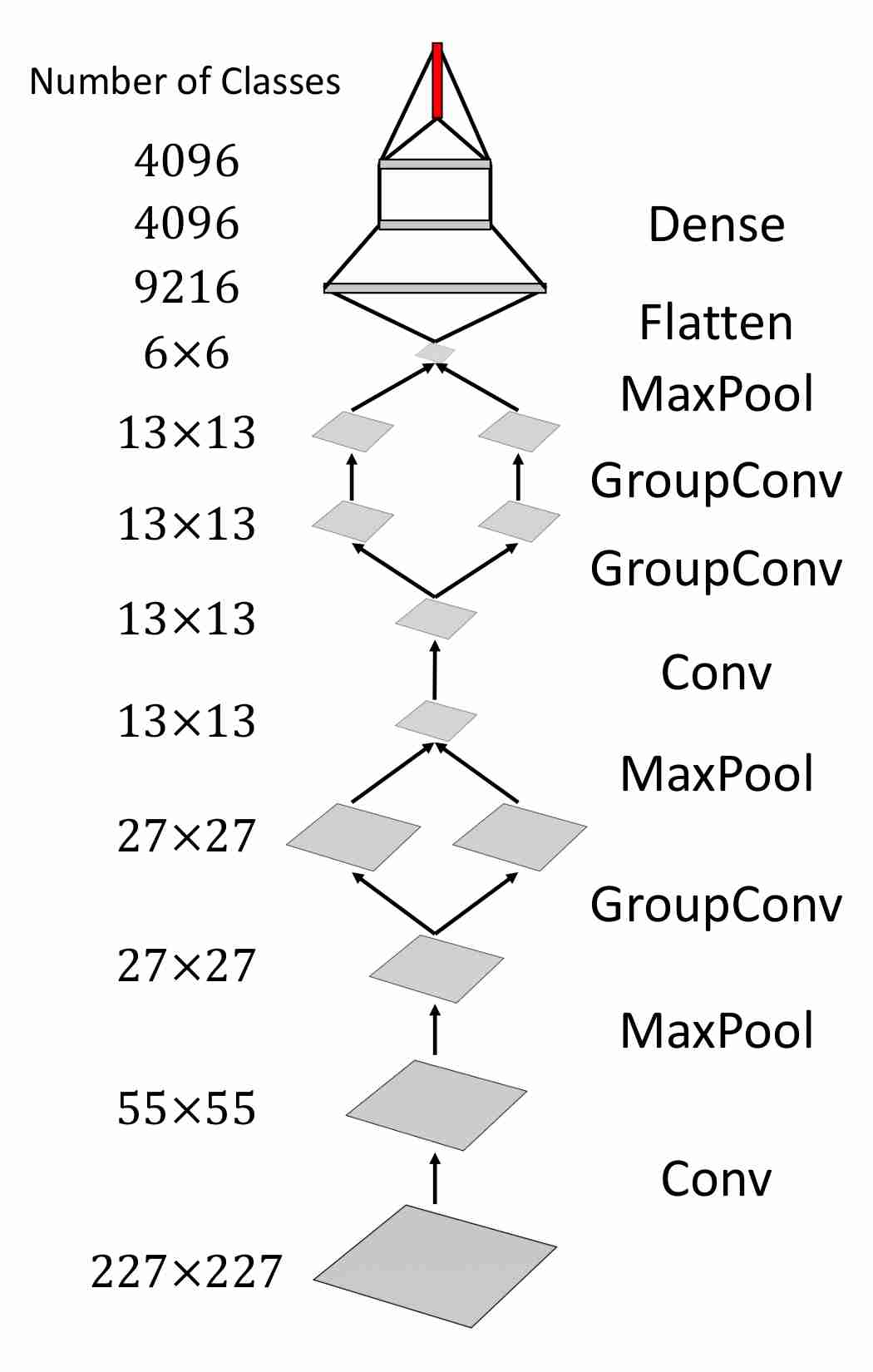}
              \captionsetup{width=.95\linewidth}
              \caption{AlexNet architecture. The group convolutional layer breaks a feature bank into two so that they can fit into multiple GPUs.}
                \label{fig_DL:alexnet}
         \end{subfigure}
          \begin{subfigure}[t]{0.605\linewidth}
                \centering
                \includegraphics[width = \linewidth, trim=0in 0in 0in 0in, clip]{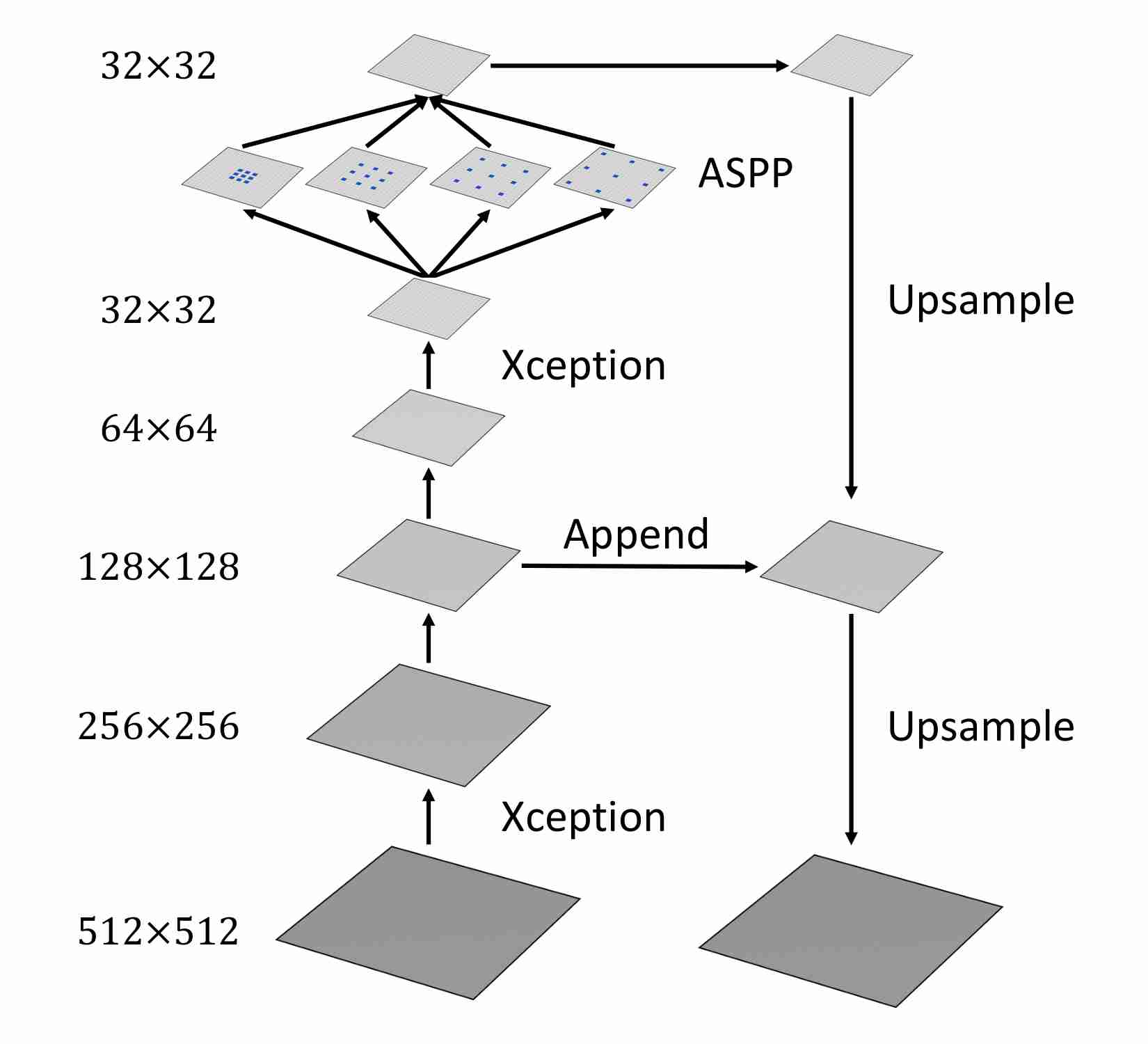}
                \captionsetup{width=.95\linewidth}
              \caption{DeepLabv3+ architecture. The ASPP uses different dilation rates parallelly in filter banks to change the field-of-view of filters, so that multi-scale information is extracted simultaneously, as depicted by the blue dots with different spaces representing multiscale configurations of pixels in the ASPP layer.}
                \label{fig_DL:deeplabv3plus}
         \end{subfigure}
         \caption{Architectures of two Deep Learning neural networks used in Section~\ref{ss_DL:cla_retri} and Section~\ref{ss_DL:sup_seg}.}
       \label{fig_DL:deep_nnet}
\end{figure}

We use an AlexNet CNN network~\cite{krizhevsky2012imagenet}~(Fig.~\ref{fig_DL:alexnet}) having five convolutional layers, some of which are followed by max-pooling layers or normalization layers, and three fully-connected layers, with {a }total of roughly $60$ million parameters and $650$ thousands neurons.

The classification network gives three outcomes simultaneously for each square patch of HR that is input to it: i) a single classified MC for each HR segment, ii) ranking of each MC within the database reflecting its similarity to the target MC of the input HR, and iii) providing an uncertainty quantification as a metric to evaluate whether the MC is a new MC not contained in the existing database. This is depicted as Step $2$ in Fig.~\ref{fig_DL:flow_chart_img_analy}. This supervised CNN model facilitates efficient integration of MC labels verified by human experts into {the database and }the classification model. For this, the model outputs some user-specified small number $K'$ {(e.g., $K'=10$) }of MCs that are most similar to the target HR (i.e., have the highest classification probability) while the remaining MCs in the database with lower classification probabilities are automatically screened out. {As part of this process, we use }evidential deep learning (EDL)~\cite{sensoy2018evidential}{ to also produce an uncertainty quantification that reflects the likelihood that the HR is a new MC not belonging to the existing database. If the uncertainty is high, an expert can then inspect micrographs for the input HR and $K'$ most similar MCs, along with their classification probabilities and uncertainty quantification, to determine the ground truth and label the HR accordingly}. 

In EDL, instead of directly modeling the probability mass function~(pmf) that the label of an input image follows{ (the number of possible labels is the number of MCs in the database)}, the neural network treats the pmf as a random vector that follows a Dirichlet distribution. The output of the neural network in EDL is thus the vector of parameters of this Dirichlet distribution, which provides not only a mean but also uncertainty estimation of prediction. More specifically, in EDL the class pmf $\bm{p}$ is assumed to follow the distribution: 
\begin{align}
\text{Dir}(\bm{p}|\bm{\alpha}) = \frac{1}{B(\bm{\alpha})} \prod_{k=1}^{K} p_k^{\alpha_k-1}
\label{eqn_DL:dirich}
\end{align}
where $\sum_{k=1}^K p_k = 1$, $\bm{\alpha}=[\alpha_1, \alpha_2, \cdots, \alpha_K]$ {are }the parameters of the distribution, and $B(\bm{\alpha})$ {is }the $K$-dimensional multinomial beta function~\cite{kotz2019continuous}. Instead of maximizing cross-entropy loss {as }in the usual classification neural network that only gives point estimation for class probabilities, EDL minimizes {the }empirical mean of the Bayes risk of a loss function for {the }pmf, which is defined for the $i ^{\mathrm{th}}$ sample as:
\begin{align}
\mathcal{L}(\bm{\Theta}; \bm{M}_i, \bm{t}_i) = \int L(\bm{p}, \bm{t}_i) \text{Dir}(\bm{p}|\bm{\alpha}_i(\bm{M}_i|\bm{\Theta})) d\bm{p}
\label{eqn_DL:bayes_risk}
\end{align}
where {$\bm{M}_i$ }and {$\bm{t}_i$ }are the input image and one-hot encoding label vector of the $i ^{\mathrm{th}}$ sample, with all parameters of {the }neural network denoted as $\bm{\Theta}$. The loss function {$L(\bm{p}, \bm{t}_i)$ }we used here is mean-squared-error~(MSE) with exponential activation function{,} as recommended in \cite{sensoy2018evidential}. The estimated class probabilities $\hat{\bm{p}}_i$ and uncertainty $\hat{u}_i$ are then calculated as $\hat{p}_{i,k} = {\alpha_{i,k}}/{\sum_{k=1}^K \alpha_{i,k}}$ and $\hat{u}_i={K}/{\sum_{k=1}^K \alpha_{i,k}}$. If an extracted HR indeed belongs to an existing MC, the uncertainty-aware classification network {ideally }gives low uncertainty; otherwise, the value of classification uncertainty from EDL should be high, indicating that the HR belongs to a new MC. {To apply EDL, the last softmax layer of the AlexNet is replaced with an exponential activation layer and the loss function contains the MSE loss term with Kullback-Leibler divergence penalization term. }

\subsection{Step $3$: Micrograph segmentation using a supervised segmentation network}
\label{ss_DL:sup_seg}
We also use a supervised segmentation network to improve the segmentation results (relative to the initial unsupervised segmentation in Step $1$) of multiphase materials, which also generates pixel-wise classification labels for MCs. The input to this segmentation network is a micrograph {that could be the same micrograph input in Step $1$ }and the output is an image of the same spatial size as the input micrograph with each pixel replaced with the MC label to which the microstructure at that pixel belongs. {This step reaches a pixel-wise accuracy higher than the unsupervised segmentation method in Step $1$ and also has the capabilities of segmentation near borders of micrographs and faster segmentation of new multiphase micrographs after the model is trained (e.g., compare Figs.~\ref{fig_DL:score_unsupervised_bgm} and~\ref{fig_DL:deeplab_seg_exp} in the later examples).} To train this model, we can either use i) data sets with full pixel-wise annotations which are usually costly; or ii) a database of homogeneous microstructures obtained in Step $2$ for which pixel-wise annotations are automatically obtained and from which {additional }multiphase micrographs with pixel-wise annotations can be easily generated to improve the model training, as described below.

In this study, a DeepLabv3+~\cite{chen2018encoder} neural network is trained to segment micrographs into HRs according to their MCs. DeepLabv3+ (Fig.~\ref{fig_DL:deeplabv3plus}) contains $\sim 41$ million parameters with $293$ layers and combines advantages of both the Atrous Spatial Pyramid Pooling~(ASPP) and the encoder-decoder structure to encode multi-scale contextual information and capture details near boundaries between different MCs. DeepLabv3+ uses DeepLabv3~\cite{chen2017deeplab} as the encoder, which gradually shrinks the spatial size of features and extracts higher-level information. The output of the feature extractor is usually $16$ times smaller than the input size, but other shrink ratios are also possible. The decoder then gradually upsamples the learned features to the same resolution as inputs. One important part of the architecture is a skip connection, appearing in almost all popular segmentation networks~(i.e., Unet~\cite{ronneberger2015u}, FCN~\cite{long2015fully}, etc.): one of middle layers of the decoder is appended along the dimension of channels with the corresponding features of the same spatial size in the encoder, before being upsampled to the same size as the input images~(see the arrow labeled ``Append" in Fig.~\ref{fig_DL:deeplabv3plus}). This skip connection also stabilizes the training and refines detailed information. In the encoder part, DeepLabv3 uses {a }modified Xception model~\cite{chollet2017xception} and ASPP model. The Xception model was further modified~\cite{qi2017deformable} in DeepLabv3 so that all max-pooling layers are replaced by a depthwise separable convolution~(DSC) layer. The DSC layer extracts multi-scale contextual information, which is crucial for capturing MC patterns at different length-scales of an MC and also demonstrates robustness against micrograph scaling and noise. To reduce the computation cost while preserving a high segmentation accuracy, we reduce the number of some stacked components. More specifically, we reduce the number of stacked Xception layers from $20$ to $8$ and retain the ASPP model. The input size is reduced from $512 \times 512$ to $256 \times 256$. The simplified DeepLabv3+ has only $\sim 1.4$ million parameters, instead of the original $41$ millions.

To generate multiphase micrographs for training the supervised segmentation network, homogeneous microstructure images are randomly selected from a database like Brodatz data sets~\cite{brodatz1966textures, efros1999texture, randen1999filtering} or the database from Steps $1$-$2$ and then randomly cropped, rotated, flipped, scaled, and pasted together as depicted in Step $3$ of Fig.~\ref{fig_DL:flow_chart_img_analy}. The boundaries of these micrograph patches can be either curved or straight lines to mimic realistic boundaries in micrographs. Then, these artificially generated multiphase micrographs and their labels are grouped as mini-batches to be used in the SGD algorithm to minimize the loss function~(\ref{eqn_DL:xentropy}) below. To accelerate the training process in all of our later examples, we used these generated data to pre-train the network, and then fine-tuned the training on real micrographs or the database with labels from Steps $1$-$2$.

To train the supervised segmentation network, we use cross-entropy classification error as the loss function and account for class imbalance of different MCs. Denote a micrograph {with $m$ pixels }and its pixel-wise {MC }labels by $(\bm {X}, \bm {Y})$ where $\bm {X} = [X_1,X_2,\cdots,X_m]$ are the pixels, and $\bm {Y} = [Y_1,Y_2,\cdots,Y_m]$($Y_i \in \{1,2,\cdots,k\}$), where $k$ denotes the total number of MCs. The loss function for this micrograph is
\begin{align}
l(\bm { \Theta}; \bm {X}, \bm {Y}) = \sum _{i=1} ^{m} \sum _{j=1} ^{k} w _{j} ^{ 0.5} I _{(Y_i=j)} \log p_{i,j}(\bm {X}; \bm { \Theta})
\label{eqn_DL:xentropy} 
\end{align}
where $\{w_j\}_{j=1} ^{k}$ are class balancing weights {that are }inversely proportional to the frequency of each MC in the training data; $I _{(Y_i=j)}$ is the indicator function that the true pixel label $Y_i$ is $j$; $p_{i,j}(\bm {X}; \bm{\Theta})$ is the predicted probability {that $Y_i = j$, and $\bm{\Theta}$ denotes }all parameters of{ the segmentation network}.

\subsection{Iteration over Steps $1$-$3$: Characterizing new micrographs and improving accuracy}
\label{ss_DL:iter_chara}
Steps $1$-$3$ can be applied to either i) a single time to a fixed set of micrographs {user knows}, or ii) iteratively, to multiple sets of micrographs, when new sets are collected over time. Most existing approaches cannot be applied to the situation ii), which is more consistent with the new material discovery/design process and allows previously learned information to be incorporated at each iteration to improve the computational efficiency and accuracy of the classification/segmentation. To make the learning loop more efficient, models in Steps $2$ and $3$ can be trained more quickly using transfer learning, as opposed to training from scratch every time micrographs of {a }new MC are obtained. For example, if the models are currently trained on a database of $k$ MCs and a micrograph representing a new $(k+1)^{\mathrm{st}}$ MC is obtained, the training can initiate with the current network weights; then in the last layer the $k$ output channels for the $k$ existing classes can be appended with an additional channel with random initial weights to initiate the new training. We show later that this performs consistently better than training-from-scratch. {In some applications, the user may be reasonably sure that the MCs that are present in a micrograph fall into a smaller subset of specified MCs. Our framework could still be applied in this case, but the classification accuracy could likely be improved if information on the narrower subset of known MCs is incorporated. This could be accomplished via the user specifying prior probabilities on the MCs in the database to indicate that the MCs in the new micrograph must belong to only a subset of the classes (e.g., by specifying a prior probability of zero that the MCs belong to any classes other than the known subset). Our framework would then produce nonzero posterior class probabilities for only the MCs in the specified subset. To guard against the possibility that the prior information is incorrect and the MC in question belongs to a class outside the specified subset, the evidential deep learning component of Step $2$ would indicate whether the MC is likely to belong to a different class.}

\section{Data sets and experimental results}
\label{s_DL:data_set_exp}
In this section, data sets and experiment results for Steps $1$-$3$ are presented in order to demonstrate each step and the effectiveness of the overall framework.

\subsection{Step $1$: Unsupervised segmentation on Brodatz, PMMA, and dual-phase steel data sets}
\label{ss_DL:nd_score_res}
\begin{figure}[!htbp]
       \centering
       \begin{subfigure}[t]{.385\linewidth} 
              \centering
              \includegraphics[width=\textwidth, trim=0in 0.0in 0in 0in, clip]{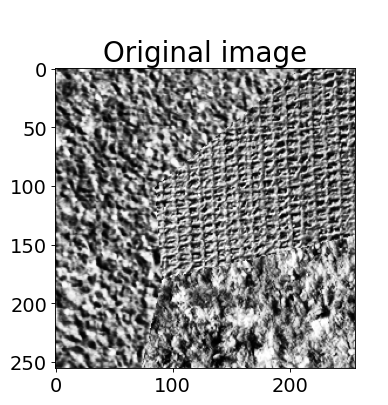}
       \caption{The original micrograph.}
       \label{fig_DL:score_unsupervised_bgm_orig}
       \end{subfigure}
       \begin{subfigure}[t]{.35\linewidth} 
              \centering
              \includegraphics[width=\textwidth, trim=0in 0.0in 0in 0in, clip]{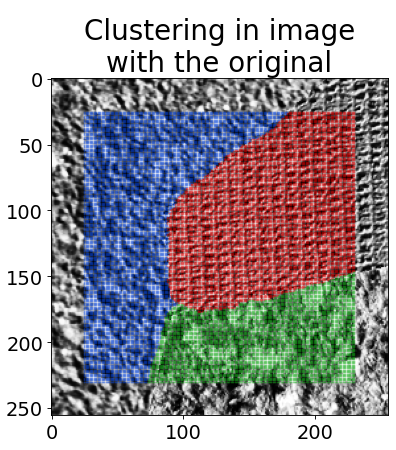}
       \caption{Unsupervised segmentation with correct number of MCs.}
       \label{fig_DL:score_unsupervised_bgm_cor_mcs}
       \end{subfigure}
       \begin{subfigure}[t]{.8\linewidth} 
              \hspace{0.45cm}
              \includegraphics[width=\textwidth]{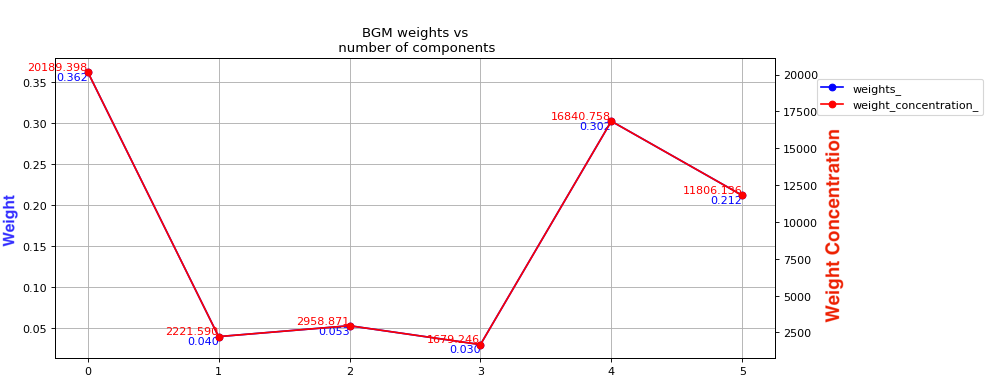}
              \caption{Posterior weights of segments in BGM applied to the original micrograph.}
              \label{fig_DL:score_unsupervised_bgm_posterior}
       \end{subfigure}
       \begin{subfigure}[t]{.65\linewidth} 
       \centering
            \includegraphics[width=\textwidth, trim=0.1in -0.2in -0.2in 0in, clip]{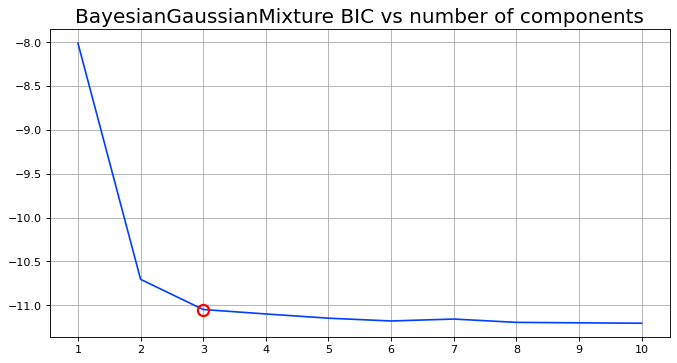}
       \caption{BIC plot for BGM applied to the original micrograph.}
       \label{fig_DL:score_unsupervised_bgm_aic_bic}
       \end{subfigure}
       \caption{Results of the BMG method of Step $1$ applied to the Brodatz data set. (a) A three-phase micrograph, created from the Brodatz data. (b) Unsupervised segmentation results for the micrograph. (c) BGM posterior weights when six clusters are used in the model, indicating that the number of HRs is $3$ (since only $3$ segments have substantial weights). (d) BIC plot for the BGM model with number of clusters varying from $1$ to $10$, which also indicates that the number of HRs is $3$ (the elbow in the plot is at $3$, beyond which the BIC decreases more slowly). The AIC generates similar plot.}
\label{fig_DL:score_unsupervised_bgm}
\end{figure}

The first data set is Brodatz microstructure, an example of which is in Fig.~\ref{fig_DL:score_unsupervised_bgm}. Each image in the database consists of only one MC, so that pixel-wise labels for each image are automatically available. With those homogeneous MCs, multiphase micrographs and their pixel-wise annotations can be generated as described in Section~\ref{ss_DL:sup_seg}. This type of generated data has been used as a benchmark to pre-train and evaluate machine learning models related to texture analysis~\cite{karabaug2019texture,todorovic2009texel,tivive2006texture,sagiv2006integrated,karoui2006region,you1993classification}, and we have found it helpful for pre-training our segmentation/classification models due to the visual similarity of the Brodatz images to many material micrographs.

\begin{figure}[!htbp]
       \captionsetup[subfigure]{width=.99\linewidth,justification=centering}
       \centering
       \begin{subfigure}[t]{.6\linewidth} 
       \centering
       \includegraphics[width=\textwidth]{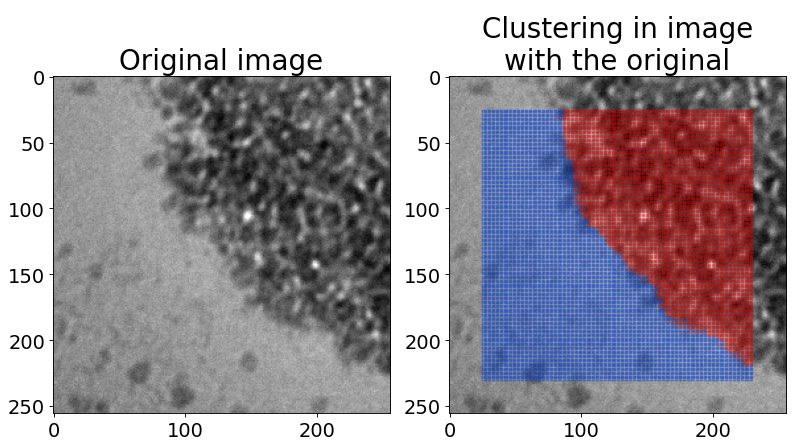}
       \caption{A TEM image of silica-PMMA.}
       \label{fig_DL:silica_PMMA_nnet_unsupervised_seg}
       \end{subfigure}
       \begin{subfigure}[t]{.99\linewidth} 
              \centering
              \includegraphics[width=\textwidth]{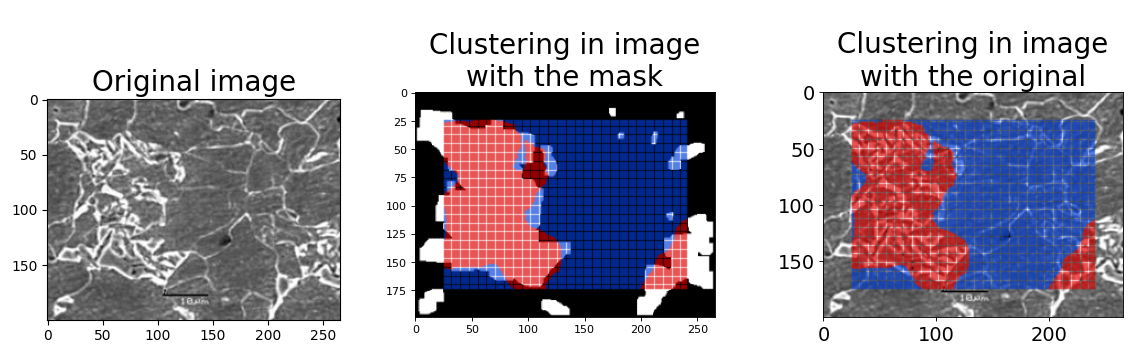}
              \caption{An SEM image of a dual-phase steel sample.}
              \label{fig_DL:dual_phase_score_unsupervised}
       \end{subfigure}
\caption{The score-based nonstationarity analysis method of Step $1$ applied to different microstructure data sets. (a) Segmentation results for a TEM micrograph of silica particles in PMMA with octyl functional modification using a linear regression model as the supervised learner. The left image is the original micrograph while the right one has the segment labels overlapped with the original micrograph. (b) Segmentation results for an SEM micrograph of dual-phase steel using a neural network with $10$ hidden layer neurons. The left image is the original micrograph. The middle and the right images are the segment labels from our method overlapped the ground truth mask and the original micrograph respectively, both of which come from~\cite{banerjee2013segmentation}. In both (a) and (b), we used parameters $l_s = 5$, $l_w=20$.}
\label{fig_DL:score_unsupervised}
\end{figure}

To perform unsupervised segmentation for this Brodatz data set, we use a linear regression model as the supervised learning model to predict individual pixel grayscale levels, for which the score vectors are computed. We also used a neural network instead of a linear regression model, but the segmentation results were nearly identical, indicating the approach is relatively robust to choice of supervised learning model~(\cite{zhang2021nonstationarity} also observed this; and we consider a neural network model for the example in Fig.~\ref{fig_DL:dual_phase_score_unsupervised}). Then, BGM is applied to the score vectors to segment the image and determine the number of MCs, which is illustrated in Fig.~\ref{fig_DL:score_unsupervised_bgm}. Fig.~\ref{fig_DL:score_unsupervised_bgm_orig} shows an example micrograph and Fig.~\ref{fig_DL:score_unsupervised_bgm_cor_mcs} shows the segmentation results overlayed on the micrograph for the BGM model with $3$ segments, which is the true number of HRs in the micrograph. The BGM method has two built-in mechanisms for determining the number of segments. One can conservatively specify the number of segments and then select the appropriate number of segments as the number of significant posterior weights (the BGM model assigns a posterior weight to each segment, which indicates the extent to which that segment is present). This is illustrated in Fig.~\ref{fig_DL:score_unsupervised_bgm_posterior} for the BGM model with 6 segments. Alternatively, one can plot BIC/AIC for the BGM versus the number of BGM segments and then select the number of segments to coincide with the ``elbow" in the plot, after which the BIC/AIC criterion decreases only gradually, which is illustrated in Fig.~\ref{fig_DL:score_unsupervised_bgm_aic_bic} for BIC (AIC plot is similar). Figs.~\ref{fig_DL:score_unsupervised_bgm_posterior} and~\ref{fig_DL:score_unsupervised_bgm_aic_bic} both correctly indicate the number of segments is $3$ for this example.

The second data set consists of TEM images of silica particles in PMMA with octyl functional modification~(Fig.~\ref{fig_DL:silica_PMMA}). Sometimes the particle dispersion is nonstationary across samples or {the particles }form large agglomerations within one sample, which can affect the physical properties of the materials~(e.g., the breakdown stress or dielectric constant{~\cite{dang2012fundamentals}}). Fig.~\ref{fig_DL:silica_PMMA_nnet_unsupervised_seg} shows a micrograph with a large agglomeration, and Fig.~\ref{fig_DL:dual_phase_score_unsupervised} shows the Step $1$ segmentation results, which correctly separate the agglomeration from the matrix region.

The third data set{~(Figs.~\ref{fig_DL:dual_phase} and~\ref{fig_DL:dual_phase_score_unsupervised}) }is an SEM image of dual-phase steel consisting of ferrite matrix and martensite in the form of islands. Segmenting the two phases is challenging because the ferrite and martensite phases share some similar features and because the martensite phases are not connected in this micrograph. Fig.~\ref{fig_DL:dual_phase_score_unsupervised} also shows our segmentation results, which agree quite closely with the ground truth MC labels. The ground truth for this micrograph was available from~\cite{banerjee2013segmentation}.

\subsection{Step $2$: {MC }Classification and ranking of predicted MCs using a CNN on Kylberg data set}
\label{ss_DL:cla_retri_res}

\begin{figure}[!htbp]
       \centering
       \includegraphics[width=\textwidth, trim=.0in 0in .0in 0in, clip]{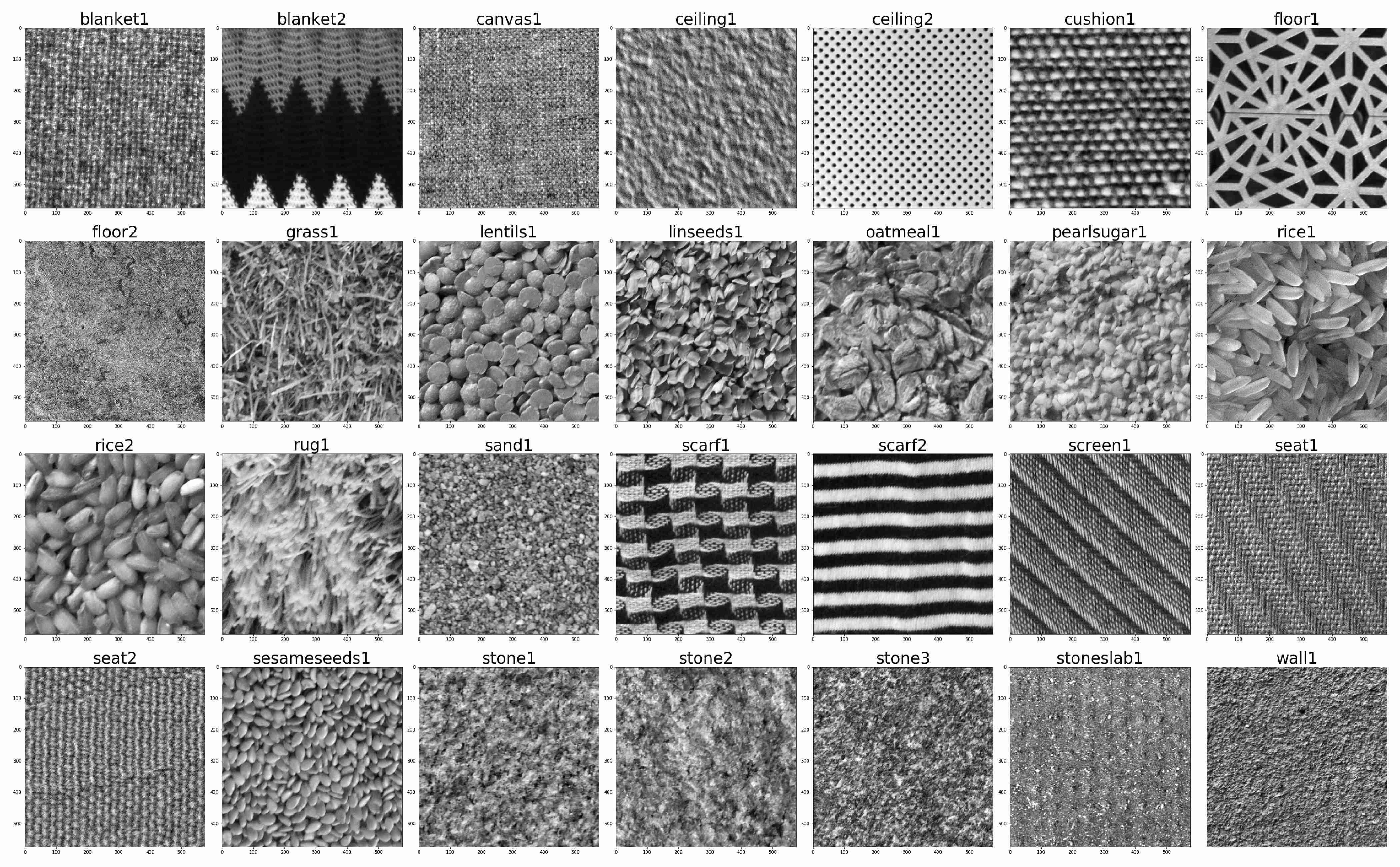}
       \caption{All micrographs in the Kylberg data set with their true label names on top of images that are used to train AlexNet classification network.}
       \label{fig_DL:kylberg_all_imgs}
\end{figure}

Here we use the Kylberg~\cite{kylberg2011kylberg} data set shown in Fig.\ref{fig_DL:kylberg_all_imgs}, which contains grayscale images of $28$ different MCs that are of sufficiently high quality\footnote{Those images are all taken under the same lighting, direction, and distance conditions.} to allow training an accurate classification model. In the Kylberg data, for each of the $28$ MCs, $4$ images were taken at different positions on a sample, and each of these images are subsequently divided into $40$ non-overlapping patches of size $576\times 576$ pixels{~\cite{kylberg2011kylberg}}. Each patch has $12$ rotated versions with in-plane rotation angles varying from $0$ to $360$ in $30$ degree increments. This augmentation increases the sample size to $53,760$ with $1,920$ image patches for each MC.

We use $80\%$, $10\%$, and $10\%$ of the data for training, validation, and testing, respectively. Images are in grayscale with $256$ levels and are not normalized (normalizing the range of pixels), but the mean values of channels from the ILSVRC data set~\cite{russakovsky2015imagenet} are subtracted from the images, according to the pre-processing procedure of pre-training AlexNet with the ILSVRC data set\footnote{http://www.cs.toronto.edu/\texttildelow{}guerzhoy/tf\_alexnet/}. The Adam optimizer~\cite{kingma2014adam} is used with default hyper-parameters.

\begin{figure}[!htbp]
       \captionsetup[subfigure]{width=\linewidth}
       \centering
       \begin{subfigure}[t]{.325\linewidth}
              \centering
              \includegraphics[width=\textwidth, trim=.0in 0in .0in 0in, clip]{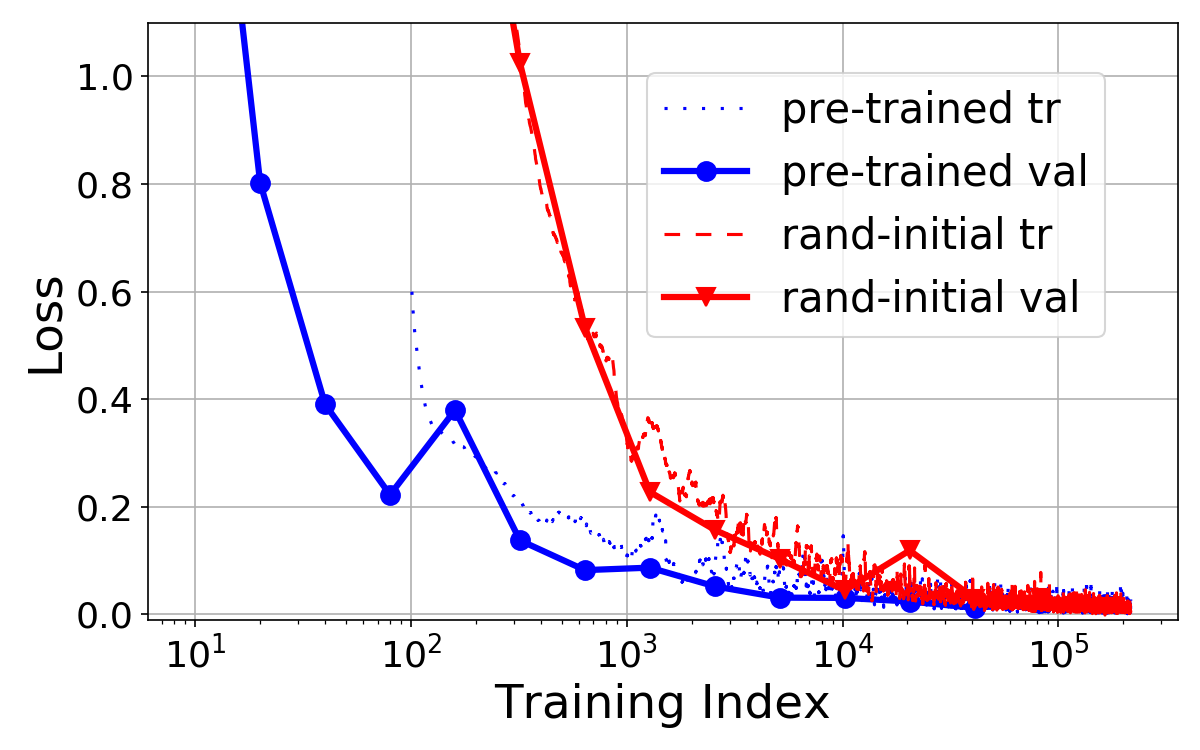}
              \caption{Loss.}
              \label{supp_fig_DL:kylber_loss}
       \end{subfigure}
       \begin{subfigure}[t]{.325\linewidth}
              \centering
              \includegraphics[width=\textwidth, trim=.0in 0in .0in 0in, clip]{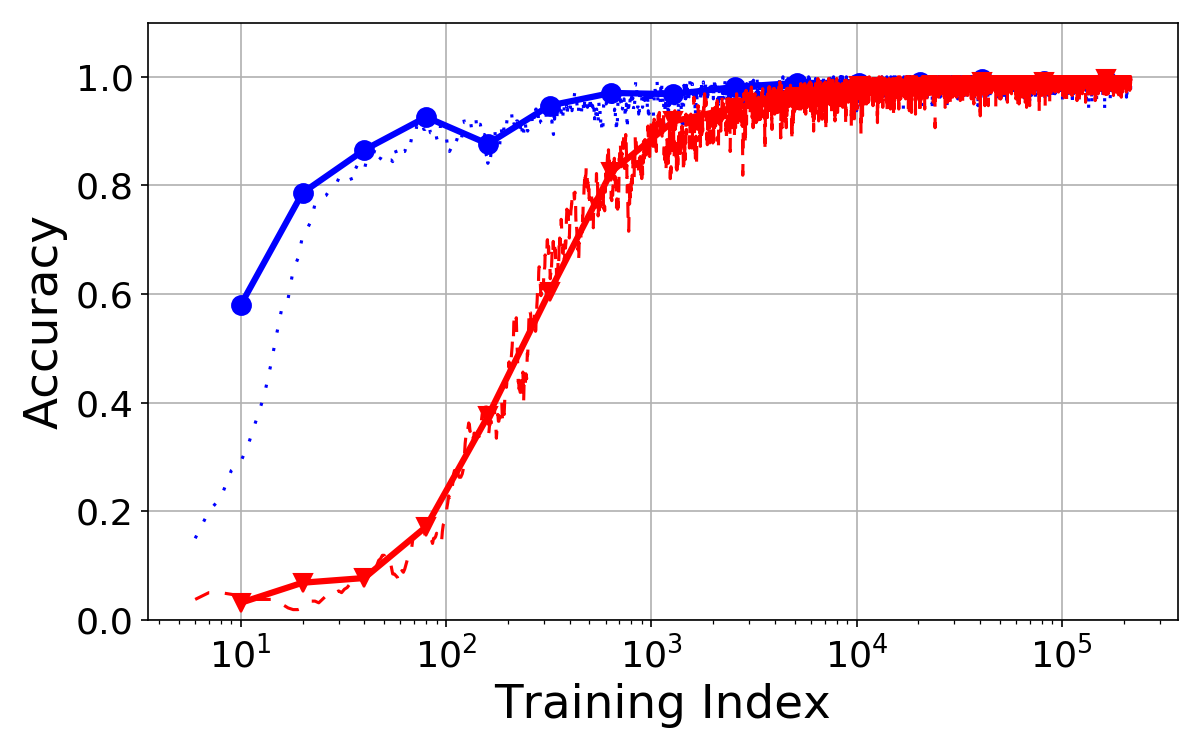}
              \caption{Accuracy.}
              \label{supp_fig_DL:kylber_acc}
       \end{subfigure}
       \begin{subfigure}[t]{.325\linewidth}
              \centering
              \includegraphics[width=\textwidth, trim=.0in 0in .0in 0in, clip]{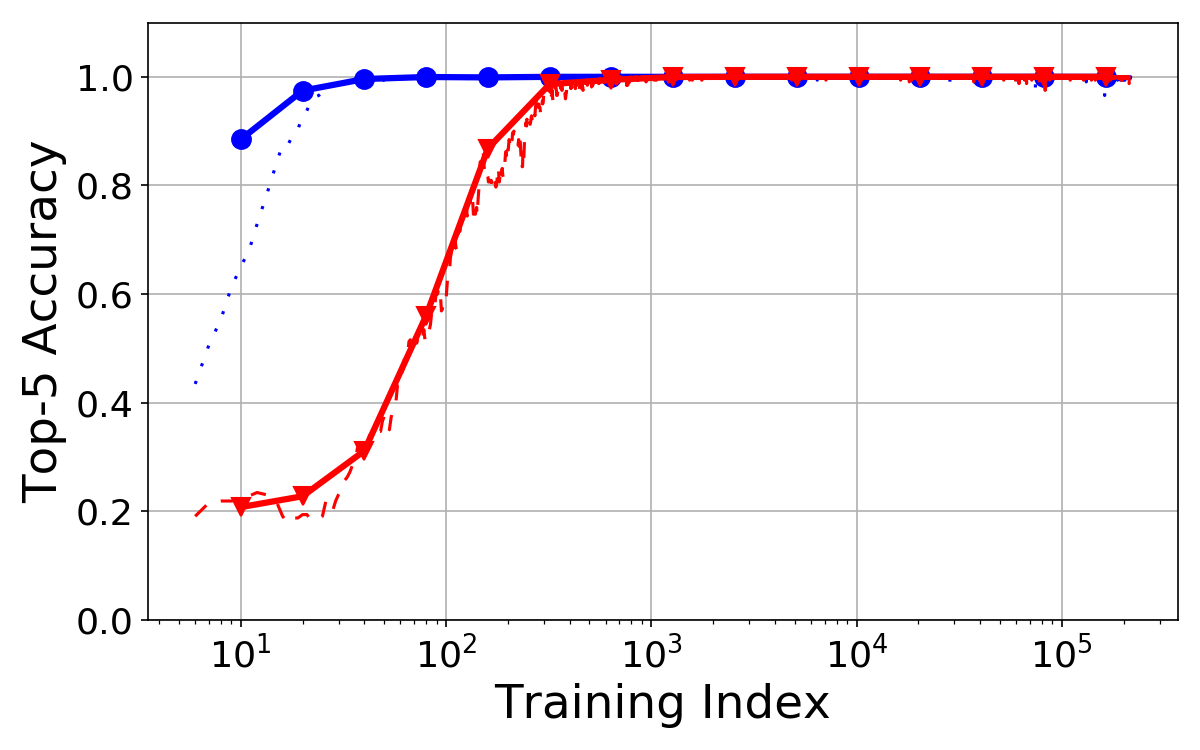}
              \caption{Top-$5$ accuracy.}
              \label{supp_fig_DL:kylber_top_5_acc}
       \end{subfigure}
       \caption{Training and validation metrics for the Step $2$ classification and ranking for the Kylberg data set demonstrating the substantial benefit of pre-training.}
       \label{fig_DL:kylberg_cla_metrics}
\end{figure}

Training such a network {from scratch }on a new data set of micrographs is time-consuming, especially when the number of MCs is large. Transfer learning techniques can significantly accelerate the training process, even though the pre-trained weights are based on the ILSVRC data set, which contains approximately $1.2$ million images representing $1000$ object categories, none of which includes microstructures~\cite{decost2017exploring} of materials. The benefit of pre-training is evident from Fig.~\ref{fig_DL:kylberg_cla_metrics}, which shows the training and validation loss, accuracy, and top-$5$ accuracy with and without pre-training. With or without pre-training the validation accuracy and top-$5$ accuracy both converge to close to $100\%$, although the pre-trained version reaches high accuracy much faster.

To demonstrate the effectiveness of {the EDL }uncertainty quantification for identifying new MCs not in the existing database in Step $2$, we held out $4$ of the $28$ MCs to serve as new ones. The validation accuracy and top-$5$ accuracy for the model trained on the micrographs for the $24$ included MCs were $98.48\%$ and $100\%$, respectively. The average EDL uncertainty for the validation data (which contained only the $24$ existing MCs) was $24.58\%$. In contrast, when the model was used to classify the micrographs for the $4$ new MCs, the average EDL uncertainty increased to $61.25\%$, which indicates that the EDL uncertainty can be effectively used to flag new MCs.

\subsection{Step $3$: Segmentation using DeepLabv3+ on Brodatz and materials data sets}
\label{ss_DL:sup_seg_res}
\begin{figure}[!htbp]
\centering
\captionsetup[subfigure]{width=1.0\linewidth}
\begin{subfigure}[t]{.49\linewidth}
         \centering
         \includegraphics[width=\textwidth, trim=.0in 0in .0in 0in, clip]{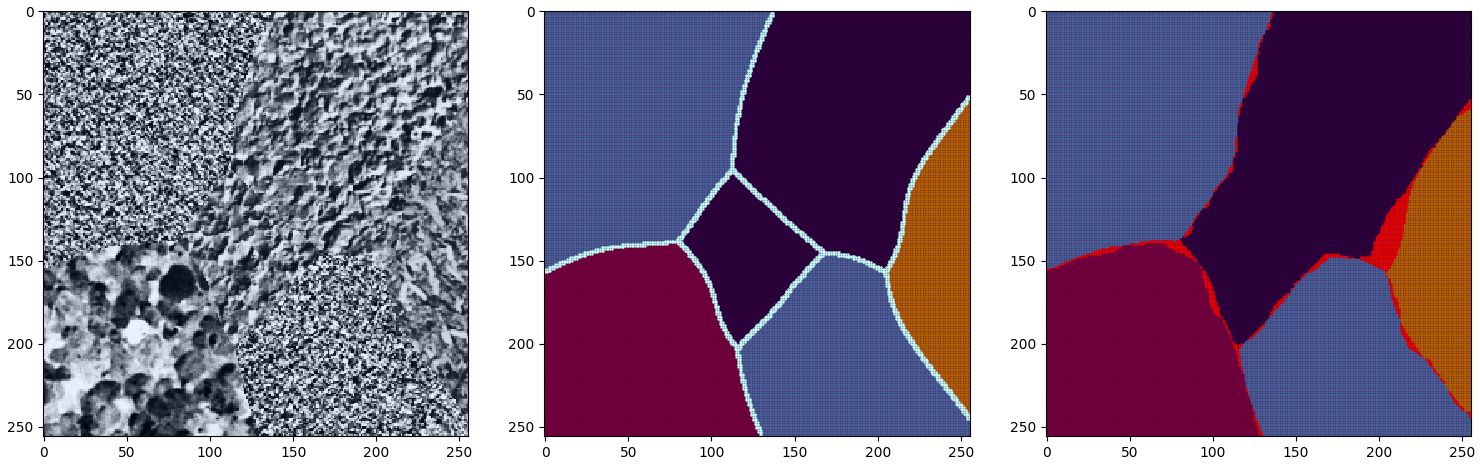}
         \caption{A Brodatz collage. }
         \label{fig_DL:brodatz_seg_exp}
  \end{subfigure}
  \begin{subfigure}[t]{.49\linewidth}
         \centering
         \includegraphics[width=\textwidth, trim=.0in .0in .0in .0in, clip]{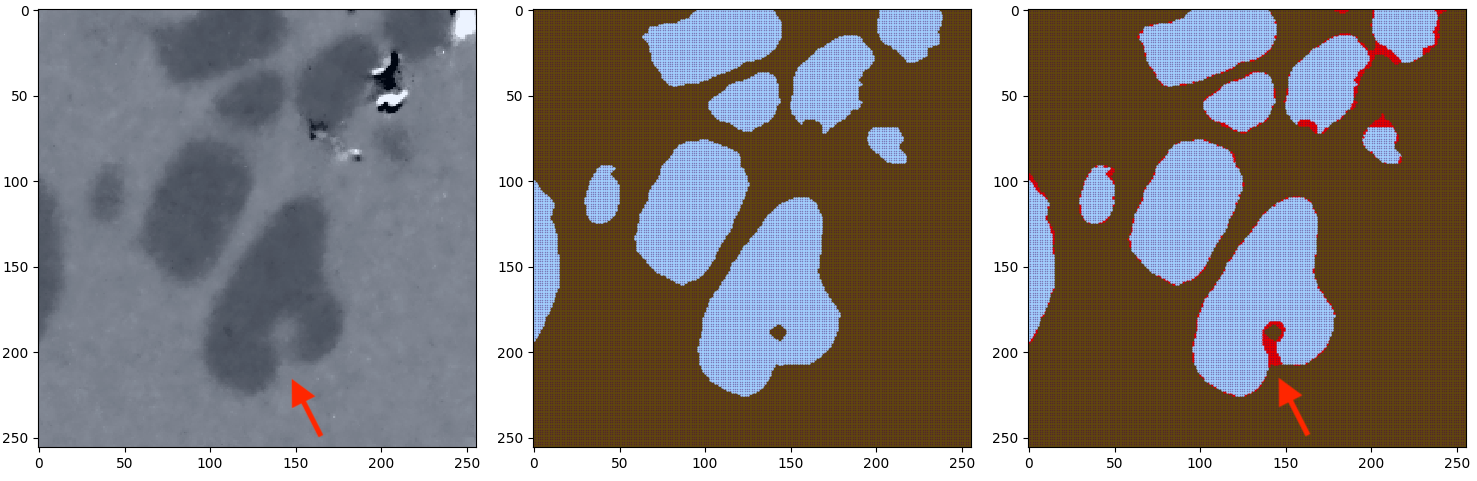}
         \caption{An XCT Al-Zn micrograph.}
         \label{fig_DL:xct_seg_exp}
  \end{subfigure}
  \begin{subfigure}[t]{.49\linewidth}
         \centering
         \includegraphics[width=\textwidth, trim=.0in .0in .0in .0in, clip]{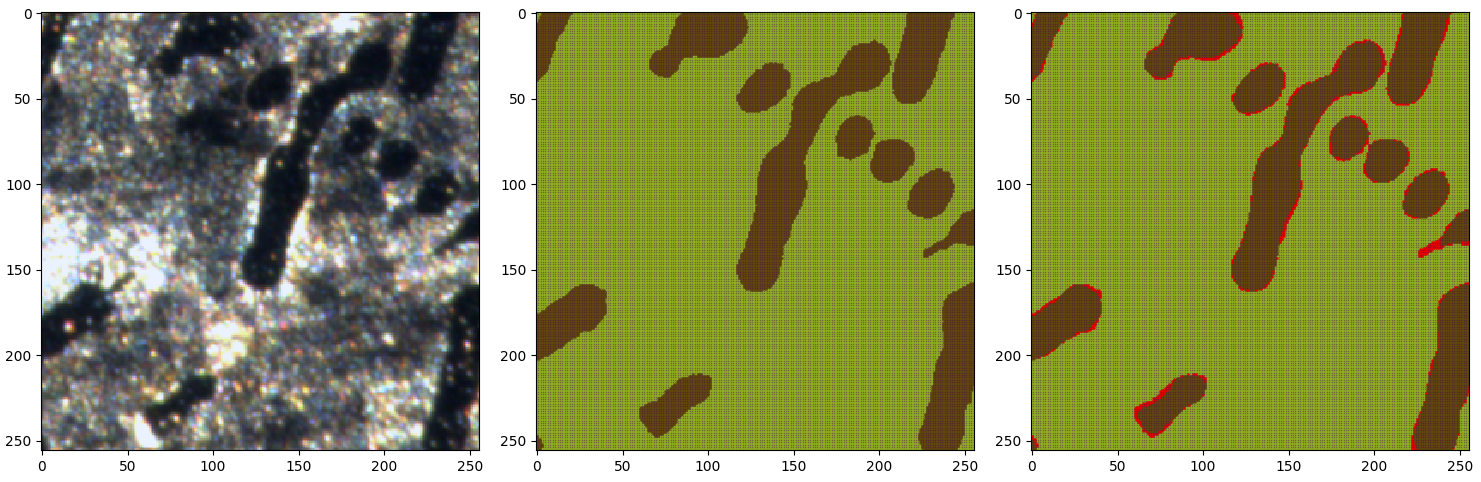}
         \caption{An SS Pb-Sn micrograph.}
         \label{fig_DL:ss_seg_exp}
  \end{subfigure}
    \begin{subfigure}[t]{.49\linewidth}
         \centering
         \includegraphics[width=\textwidth, trim=.0in .0in .0in .0in, clip]{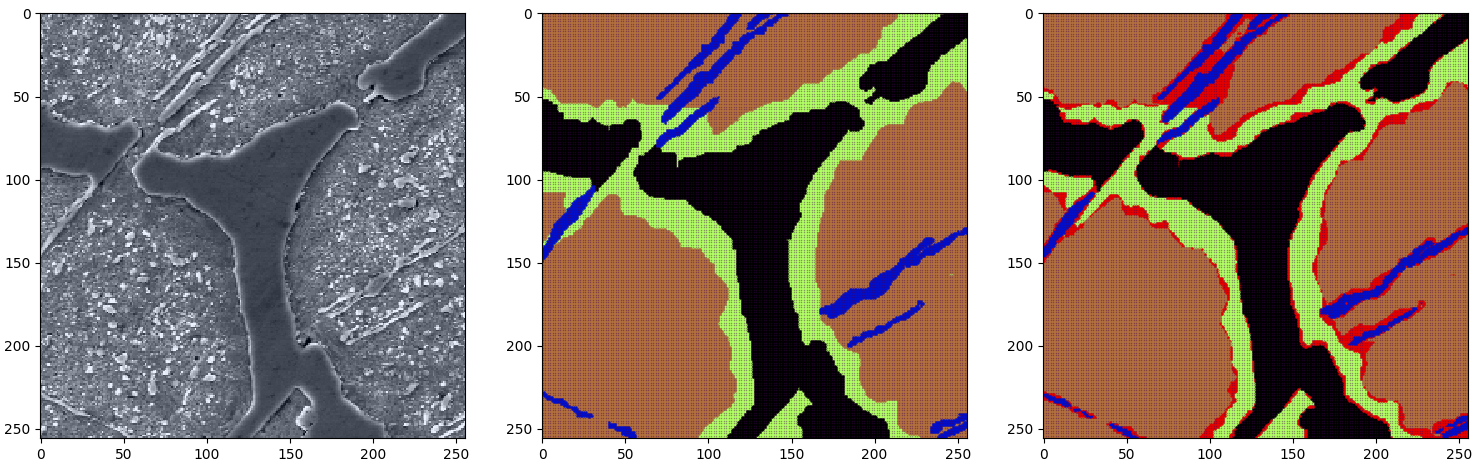}
         \caption{An SEM UHCS micrograph.}
         \label{fig_DL:uhcs_seg_exp}
  \end{subfigure}
\caption{Test results for the Step $3$ supervised segmentation of the Brodatz and real materials data sets. For each subfigure, the left image is the micrograph to be segmented; the middle image is the mask showing the ground truth coded with different colors for different MCs (the boundaries between patches in (a) are colored differently); the right image is the predicted labels, with the same color scheme with red indicating misclassified pixels. (a) $4$ MCs are different colors. (b) Al and Zn MCs are brown and light blue, respectively. (c) Pb and Sn MCs are green and brown, respectively. (d) ferritic, proeutectoid, spheroidite, and Widmanst\"atten MCs are green, black, brown, and blue, respectively.
}
\label{fig_DL:deeplab_seg_exp}
\end{figure}

We applied the supervised segmentation to four data sets: Brodatz, two dendrite growth{ data sets}~(one a Al-Zn alloy with X-ray computed tomography micrographs~(Fig.~\ref{fig_DL:xct} or~\ref{fig_DL:xct_seg_exp}) and the other a Pb-Sn alloy with serial sectioning microscopy micrographs~(Fig.~\ref{fig_DL:ss} or~\ref{fig_DL:ss_seg_exp}))~\cite{stan2020optimizing}, and UHCS{~(Fig.~\ref{fig_DL:uhcs_singlephase},~\ref{fig_DL:uhcs_multiphase}, or~\ref{fig_DL:uhcs_seg_exp})}. We train a segmentation model for each data set. The Brodatz data set is generated according to the data-augmentation process described in Section~\ref{ss_DL:sup_seg}. Because pixel-wise annotations are obtained automatically, collages of $5$ MCs with random and arbitrary boundaries are generated along with labels of MCs as ground truth. The collages as well as boundaries between different HRs are randomly generated for training, validation, and testing data splits so that they are not the same across different data splits. The sample sizes of augmented training, validation, and testing data splits are $20,000$, $5,000$, and $10,000$, respectively. For other real materials data sets, only rotation and flipping are used to augment the data splits. The Al-Zn alloy data set contains $42$ original micrographs with resolution $852 \times 852$. Smaller micrograph patches of size $256 \times 256$ are generated from each original micrograph to meet the input dimension of the reduced-size segmentation network described in Section~\ref{ss_DL:sup_seg}. The training, validation, and testing data splits are generated from $30$, $10$, and $2$ original micrographs through this patch-generation process, respectively. The sample sizes of augmented training, validation, and testing data splits are $150,000$, $21,000$, and $10,500$, respectively. The Pb-Sn alloy data set contains $9$ original micrographs with resolution $1689 \times 985$. The training, validation, and testing data splits are generated from $6$, $2$, and $1$ original micrographs using the same patch-generation process. The sample sizes of augmented training, validation, and testing data splits are $370,000$, $36,000$, and $14,000$, respectively. The UHCS data set contains $24$ original micrographs as described in~\cite{decost2019high}, each with the resolution $645 \times 522$ but with different scale bars. For this data set, we evaluate the cross-validation accuracy with each fold generated by a group of $4$ original micrographs (in total, we have $6$ such groups) through the same patch-generation process. The augmented sample sizes of a training split and a validation fold are $140,000$ and $3,500$, respectively, and we cycle through all $6$ folds. Because each micrograph is at a different scale, we rescaled all micrographs so that pixels correspond to the same physical length.

As {seen }in the right plots of the subfigures in Fig.~\ref{fig_DL:deeplab_seg_exp}, the predicted MC labels for the test cases agree quite closely with the true labels in the middle plots, and only a small portion of pixels near the boundary regions are mis-segmented (boundaries are obviously more difficult to classify and sometimes ambiguous in real micrographs). For the Brodatz data set in Fig.~\ref{fig_DL:brodatz_seg_exp}, the overall (pixel-wise) test accuracy is $97.8\%$ with all MCs having true positive rates larger than $96\%$; for the Al-Zn micrographs in Fig.~\ref{fig_DL:xct_seg_exp} the overall test accuracy is $99.51\%$, and the true positive rate for the Al and Zn are $99.41\%$ and $99.61\%$; for Pb-Sn micrographs, the overall test accuracy is $97.58\%$, and the true positive rate for Pb and Sn are $97.83\%$ and $96.70\%$; for UHCS micrographs in Fig.~\ref{fig_DL:uhcs_seg_exp}, the overall test accuracy is $93.1\%$, and the true positive rate for ferritic, proeutectoid, spheroidite, and Widmanst\"atten are $90.7\%$, $94.2\%$, $95.1\%$, and $89.6\%$ respectively.

In addition to achieving high accuracy, our supervised segmentation network has other desirable features. One issue often encountered in real applications is that the scale of micrographs may differ from sample to sample. For example, different SEM images of the same microstructure may have somewhat different magnifications. To test the impact of small scaling variations on micrograph segmentation, for each patch in a collage we randomly choose the scale of MCs to lie within the interval $[1, 1.1]$, enlarge them accordingly, and then crop and paste all patches together. Our experiments show that the test accuracy~(loss) are approximately the same as without scaling. 
This is likely because the ASPP layer in DeepLabv3+ extracts multi-scale patterns in one layer and thus is robust to such small changes in scaling.

Segmenting real micrographs can be challenging because of some ambiguity in true labels. As in Fig.~\ref{fig_DL:xct_seg_exp}, there is a small island of class $1$~(Al) in one branch of the dendrites (the red arrows) and the algorithm correctly captures that, even though it misclassifies a bridge that connects the island and the rest of {the }bulk Al. From the original image, the misclassified part has much lighter intensity than other parts of class $2$~(Zn), which is likely the reason our algorithm misclassified it.

{Figs.~\ref{fig_DL:xct_seg_exp} and~\ref{fig_DL:uhcs_seg_exp} also demonstrate the flexibility and power of this framework, because some of the homogeneous segments are stretched in one direction and thin in another direction. This would present challenges for an alternative approach in which one divided the micrograph into patches, computed an NPCF for each patch, and then clustered the NPCFs. It would be difficult to determine the size of patches, since they would have to be large enough to allow the NPCF to be computed at sufficiently large correlation distances, but large patches could overlap multiple different homogeneous segments. In contrast, our framework involves a separate score vector for each individual pixel in Step $1$ and pixel-wise prediction in Step $3$, which allows homogeneous segments of any shape to be accurately identified.}

We also found that transfer learning accelerates the training of the Step $3$ models, similar to what we described under Step 2 (Fig.~\ref{fig_DL:kylberg_cla_metrics}). In particular, we found that pre-training on the Brodatz data before training on real material data substantially accelerated the training, relative to using random initial weights when training on the real data (results omitted for brevity), in spite of the Brodatz data having its own stochastic nature that may or may not be similar to real material micrographs.

\subsection{Iteration over Steps $1$-$3$: Iteratively characterizing new samples on Brodatz data set}
\label{ss_DL:iteri_res}

We {now }use the Brodatz data set to simulate the scenario where Steps $1$-$3$ are iteratively repeated to identify new MCs and integrate them into the database, because the number of MCs is easily controlled. We begin with $4$ MCs and increase their number by one at each iteration until the database contains 22 MCs. 

\begin{figure}[!htbp]
\centering
\begin{subfigure}[t]{0.425\linewidth}
              \centering
              \includegraphics[width=\textwidth, trim=.0in 0in 3.5in 0in, clip]{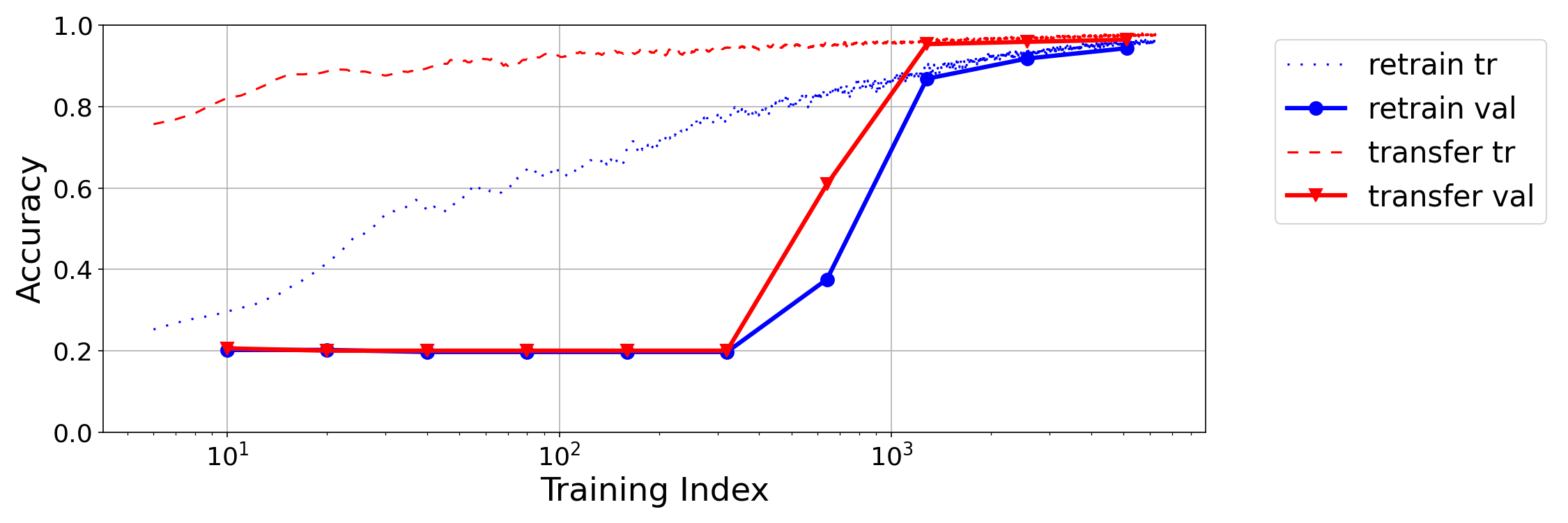}
              \caption{$5$ classes to differentiate.}
              \label{fig_DL:incr_learning_5_cla}
       \end{subfigure}
       \begin{subfigure}[t]{0.55\linewidth}
              \centering
              \includegraphics[width=\textwidth, trim=.0in 0in .0in 0in, clip]{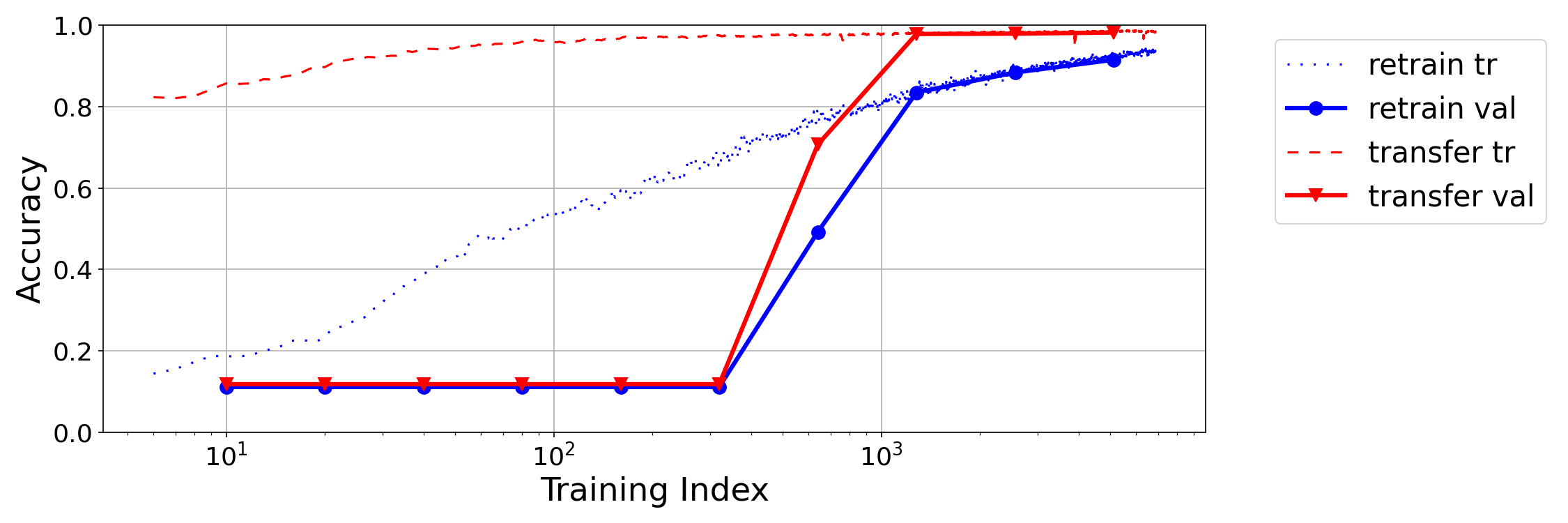}
              \caption{$9$ classes to differentiate.}
              \label{fig_DL:incr_learning_9_cla}
       \end{subfigure}  
\caption{Training (tr) and validation (val) accuracy for two ways of training when iterating over Steps $1$-$3$ (results shown at iterations $k = 5$ and $k = 9$): Complete retraining with all weights randomly initialized~(``retrain''), transfer learning with only the weights of the new channel associated with the new MC randomly initialized, followed by training the entire network~(``transfer'').}
       \label{fig_DL:incr_cla_deeplab}
\end{figure}

In Sec.~\ref{ss_DL:cla_retri_res} and~\ref{ss_DL:sup_seg_res}, transfer learning was helpful for accelerating model training. Here, we demonstrate that it is also helpful when retraining the models at each iteration of Steps $1$-$3$. Suppose that a supervised segmentation model in Step $3$, denoted by $\mathcal{M}_k$, has been trained on a database with $k$ MCs. The naive approach of training the supervised segmentation model when the $(k+1)^{\mathrm{st}}$ MC is added is to retrain the model $\mathcal{M}_{k+1}$ with all weights randomly initialized. Alternatively, we can retain the parameters in the trained $\mathcal{M}_{k}$, expand the number of channels in the last layer to classify $k+1$ classes, randomly initialize only those newly added weights associated with the channel corresponding to the $(k+1)^{\mathrm{st}}$ MC, and continue training the network with old and new samples. As seen in Fig.~\ref{fig_DL:incr_cla_deeplab}, as the number of classes increases, the transfer learning is consistently better than complete retraining method and achieves very high accuracy when new MCs are added, as well as faster convergence.

{Regarding the wall-clock time for each step, we give example times for some data sets below. Note that the times could be further improved if the code implementation and hardware are optimized. The algorithms were implemented on a Linux server running Red Hat Enterprise Linux $8.10$ (Ootpa) with an Intel(R) Xeon(R) Gold $6230$R CPU @ $2.10$GHz ($24$ cores), $256$ GB RAM, and an NVIDIA H$100$ GPU cards. For the Step $1$ unsupervised segmentation results shown in Fig.~\ref{fig_DL:score_unsupervised}, the time for Fig.~\ref{fig_DL:silica_PMMA_nnet_unsupervised_seg} using the linear model is about $3$ minutes for a micrograph of size $256 \times 256$ pixels, and the time for Fig.~\ref{fig_DL:dual_phase_score_unsupervised} using the neural network is about $36$ minutes for a micrograph of size $200 \times 266$ pixels. For the Step $2$ classification results for the Kylberg data set in Fig.~\ref{fig_DL:kylberg_all_imgs}, after the classification deep neural network is trained the time for classifying each MC image patch is very fast (around $1$ second). Training the classification deep neural network from scratch on the $53,760$ image patches took about $4$ hours. However, the classification model training for Step $2$ only needs to be updated periodically, and transfer learning substantially speeds up this process, as depicted in Fig.~\ref{fig_DL:kylberg_cla_metrics}. For the Step $3$ supervised segmentation results for the UHCS data set in Fig.~\ref{fig_DL:uhcs_multiphase}, after the segmentation deep neural network is trained the time for segmenting each MC image patch is also very fast (around $1-3$ seconds). Training the segmentation deep neural network from scratch on the $140,000$ augmented images is about $8$ hours. Step $3$ also only needs to be updated periodically, when new MCs are discovered and integrated into the database, and transfer learning can speed up this step too, as shown in Fig.~\ref{fig_DL:incr_cla_deeplab}.}

\section{Conclusions}
\label{s_DL:conclusion}
This study integrates unsupervised segmentation, supervised uncertainty-aware classification, and supervised segmentation to provide a comprehensive framework characterizing microstructures of multiphase materials, an increasingly important goal in materials science. This framework is designed for iterative use, with each iteration characterizing and segmenting multiphase materials and integrating newly identified MCs into a database, improving the characterization performance at each iteration. We have demonstrated the performance of each procedural step on a variety of real and synthetic micrograph examples and also demonstrated the ability to identify new MCs (through evidential deep learning) and incorporate them into the database to facilitate the characterization of a broader range of materials. We have also shown that data augmentation and transfer learning techniques can be used to enhance the performance of the framework. One limitation of the framework is that it is designed to characterize microstructures having stochastic behavior. For microstructures with more structured geometric patterns, further investigation is needed. 

\section*{CRediT authorship contribution statement}
\textbf{Kungang Zhang}: Conceptualization, Methodology, Software, Validation, Formal analysis, Investigation, Data curation, Resources, Writing - original draft, Writing - review \& editing, Visualization, Project administration, Funding acquisition. \textbf{Daniel W. Apley}: Conceptualization, Methodology, Validation, Writing - review \& editing, Supervision, Funding acquisition. \textbf{Wei Chen, Wing K. Liu, L. Catherine Brinson}: Writing - review \& editing, Supervision, Funding acquisition.

\section*{Declaration of competing interest}
The authors declare that they have no known competing financial interests or personal relationships that could have appeared to influence the work reported in this paper.

\section*{Acknowledgement}
This research work was partially supported by the Air Force Office of Scientific Research under grants FA9550-14-1-0032 and FA9550-18-1-0381, for which we express our sincere gratitude. Additionally, this work was also supported by funded resources from the Extreme Science and Engineering Discovery Environment (XSEDE)~\cite{towns2014xsede} (NSF grant ACI-1548562) and the Advanced Cyberinfrastructure Coordination Ecosystem: Services \& Support (ACCESS) program~\cite{boerner2023access} (NSF grants 2138259, 2138286, 2138307, 2137603, and 2138296). Further computational resources were provided by the Quest high-performance computing facility at Northwestern University, which is jointly supported by the Office of the Provost, the Office for Research, and Northwestern University Information Technology. The micrographs of silica particles in PMMA are kindly provided by Prof. Linda Schadler (Linda.Schadler@uvm.edu).


\bibliographystyle{Perfect}

\bibliography{ref.bib}
\end{document}